\documentclass[fleqn,usenatbib]{mnras}

\usepackage{newtxtext,newtxmath}

\usepackage[T1]{fontenc}

\DeclareRobustCommand{\VAN}[3]{#2}
\let\VANthebibliography\thebibliography
\def\thebibliography{\DeclareRobustCommand{\VAN}[3]{##3}\VANthebibliography}

\usepackage{graphicx}	
\usepackage{amsmath}	

\title[Deuterated formaldehyde in VLA1623-2417]{FAUST -- XXII. Deuteration in the VLA1623-2417 protostellar hot-corinos, cavities, and streamers.}

\author[S. Mercimek et al.]{
S. Mercimek,$^{1,2}$\thanks{E-mail: seyma.mercimek@manchester.ac.uk}
C. Codella$^{2,3}$,
L. Podio$^{2}$,
P. Caselli$^{4}$,
C. J. Chandler$^{5}$,
L. Chahine$^{3}$,
S. Ohashi$^{6}$,
G. Sabatini$^{2}$,
\newauthor
L. Loinard$^{7,8}$,
D. Johnstone$^{9,10}$,
E. Bianchi$^{2}$,
Y. Zhang$^{11}$,
M. De Simone$^{12}$,
C. Ceccarelli$^{3}$, 
N. Sakai$^{6}$,
\newauthor
S. Yamamoto$^{13}$
\\
$^{1}$ Jodrell Bank Centre for Astrophysics, UK ALMA Regional Centre Node, University of Manchester, Manchester M13 9PL, UK \\
$^{2}$ INAF, Osservatorio Astrofisico di Arcetri, Largo E. Fermi 5, I-50125, Firenze, Italy \\
$^{3}$ Univ. Grenoble Alpes, CNRS, IPAG, 38000 Grenoble, France\\
$^{4}$ Max-Planck-Institut für extraterrestrische Physik (MPE), Gießenbachstr. 1, D-85741 Garching, Germany\\
$^{5}$  National Radio Astronomy Observatory, PO Box O, Socorro, NM 87801, USA\\
$^{6}$ RIKEN Cluster for Pioneering Research, 2-1, Hirosawa, Wako-shi, Saitama 351-0198, Japan\\
$^{7}$Instituto de Radioastronomía y Astrofísica , Universidad Nacional Autónoma de México, A.P. 3-72 (Xangari), 8701, Morelia, Mexico\\
$^{8}$Instituto de Astronomía, Universidad Nacional Autónoma de México, Ciudad Universitaria, A.P. 70-264, Ciudad de México 04510, Mexico\\
$^{9}$NRC Herzberg Astronomy and Astrophysics, 5071 West Saanich Road, Victoria, BC V9E 2E7, Canada\\
$^{10}$Department of Physics and Astronomy, University of Victoria, Elliott Building, 3800 Finnerty Road, Victoria, BC V8P 5C2, Canada\\
$^{11}$Department of Astronomy, University of Virginia, Charlottesville, VA 22904, USA\\
$^{12}$European Southern Observatory, Karl-Schwarzschild-Strasse 2, 85748, Garching bei München, Germany \\
$^{13}$ SOKENDAI (The Graduate University for Advanced Studies), Shonan Village, Hayama, Kanagawa 240-0193, Japan
}

\date{Accepted XXX. Received YYY; in original form ZZZ}

\begin{document}
\label{firstpage}
\pagerange{\pageref{firstpage}--\pageref{lastpage}}
\maketitle

\begin{abstract}
The study of deuterium fractionation is a valuable tool for reconstructing our chemical history from the early prestellar stages to the formation of planets. In the context of the ALMA Large Programme FAUST, we observed formaldehyde, H$_2$CO, and its singly and doubly deuterated forms, HDCO and D$_2$CO, towards the protostellar cluster VLA1623--2417, on scales of $\sim2000-50$ au. 
Formaldehyde probes the inner envelopes of the protostars VLA1623A, B, and W, the rotating cavities opened by the VLA1623A outflow, and several streamers.
The HDCO and D$_2$CO emissions are observed towards VLA1623A, in its outflow cavities, and in one of the streamers.
We estimate the gas temperature from the  HDCO lines: T$\sim125$ K towards VLA1623A, indicating hot-corino emission, lower temperatures in the outflow cavities ($20-40$ K), and in the streamers ($\le15$ K). 
The D$_2$CO lines also trace the flattened envelope of VLA1623A, where H$_2$CO and HDCO are fainter. This may be due to D$_2$CO formation on dust grains in the cold prestellar phase, and subsequent photodesorption caused by the enhanced UV flux from two nearby B stars.
We inferred the molecular deuteration: [HDCO]/[H$_2$CO] $\sim0.16$, $\sim0.07-0.13$, and $\sim0.3$; [D$_2$CO]/[H$_2$CO] $\sim0.003$, $\sim0.05-0.13$, and $\sim0.03$ in the hot corino, in the outflow cavities, and in the streamer, respectively. 
The spatial distribution of D$_2$CO, which points to formation on dust grains, and the similar values of [HDCO]/[H$_2$CO] and [D$_2$CO]/[H$_2$CO] in the components of the system, suggest that deuterium fractionation occurs at the prestellar stage and is then inherited, mostly unaltered, in the protostellar phase.

\end{abstract}




\section{Introduction}
Observing how molecular complexity evolves in Sun-like star-forming regions is mandatory to comprehend whether the chemical composition of the protostellar stages is inherited by protoplanetary disks and, possibly, newly formed planets \citep[e.g.,][]{Caselli2012}. In this context, the deuterium-bearing species are a powerful tool to investigate the chemical evolution along the star formation process \citep[e.g.,][]{Ceccarelli2014, Ceccarelli2023}. The deuterium fraction, as observed in the early (prestellar, protostellar) stages of the star-forming process, can be much larger than the deuterium elemental abundance (D/H $\sim$ 2.5$\times$10$^{-5}$, \citet{Cooke2018}.
Significant deuteration fraction is seen in the cold ($\leq$ 10 K), dense ($\geq$ 10$^{5}$ cm$^{-3}$) prestellar cores where most of the species freeze out onto the cold surfaces of the dust grains \citep[e.g.,][]{Caselli2002, Bacmann2003, Crapsi2005, Sabatini2020}. When CO is heavily depleted, as in prestellar cores \citep[e.g.,][]{Caselli1999, Bacmann2002, Sabatini2022}, H$^{+}_{3}$ can efficiently react with HD  
via the following exothermic  reaction to form its single-deuterated form \citep[e.g.,][]{Gerlich2002,Ceccarelli2014}:

\begin{equation}
        H^{+}_{3} + HD \rightarrow H_{2}D^{+} + H_{2} + 232 K.
\end{equation}

H$_{2}$D$^{+}$ can further reacts with HD to form first D$_{2}$H$^{+}$, and then D$_{3}^{+}$ \citep[e.g.,][]{vastel2004}.
Once the protostar forms, the accretion of
envelope material, the temperature of the protostellar envelope increases, and molecules such as CO and N$_2$ are released back in the gas phase and react with the deuterated ions, producing an increase in the D/H ratio of several species \citep{Emprechtinger2009}. This applies to 
both the so-called Class 0 (age $\geq$ 10$^4$ yr),
and Class I ($\geq$ 10$^5$ yr) stages \citep{Andre1990}.


Typical examples of species where D/H shows a dramatic enhancement, up to several orders of magnitude, are water, methanol, and formaldehyde \citep[e.g.,][and references therein]{Loinard2001,vastel2004, Ceccarelli2007, Caselli2012, Persson2018, Taquet2019, Podio2024, Chahine2024}.
More specifically, the fractional deuteration of formaldehyde (H$_{2}$CO) is 
an excellent probe to investigate the chemical evolution along the star formation process, being detected from cold prestellar cores \citep[e.g., ][]{Chacon2019} to protoplanetary disks with ages
larger than 10$^5$ yr \citep[e.g., ][]{Podio2024}. 
Formaldehyde can be considered an intermediate species in the process of forming complex organic molecules. It can be formed either in the gas phase or on grains.
Gas-phase reactions start with C$^{+}$ to form CH$_{3}^{+}$ followed by oxidation \citep[e.g.,][]{roberts2000,Turner2001,Roueff2007}. 
On the other hand,  grain surface formation occurs via multiple hydrogenation of frozen CO \citep[e.g.,][and references therein]{Tielens1983, Fuchs2009, Garrod2022}. The deuterated forms of H$_{2}$CO are either formed in gas-phase or stored in the grain mantles, where they remain until the protostar is formed and the grain mantles are heated and gradually evaporated \citep[e.g.,][]{Ceccarelli2007}. 

Gas-phase formaldehyde deuterium fraction has been studied towards low-mass star-forming objects using single-dish antennas \citep[e.g.,][]{Bacmann2003, Parise2006, Bianchi2017,Mercimek2022} and more recently interferometric observations taken in the context of the ALMA programs PILS and FAUST \citep[e.g.,][]{Persson2018,Manigand2019, Evans2023, Podio2024}. 
These observations lead to D/H values $>$ 10$^{-2}$ and $>$ 10$^{-3}$ for HDCO, and D$_2$CO, respectively. In addition, what has been found so far shows no statistically significant difference in the deuteration fractions in prestellar cores, Class 0 protostars, and Class I/II disks. This suggests that the deuterated molecules in protostars and disks are inherited from the prestellar phase, with little or no reprocessing \citep{Drozdovskaya2021}. 
However, the available estimates are obtained mainly by using single-dish observations. Therefore, the inferred deuteration is an average over the various structures (hot-corinos, outflows, streamers, envelopes) covered by the beam. Interferometric observations are therefore mandatory to minimise beam dilution and disentangle the inner region (less than 100 au) around protostars from the other physical components such as envelopes, outflows, and infalling streamers.  

In this paper, we present observations of formaldehyde and its single- and doubly-deuterated isotopologues toward the prototypical VLA1623--2417 protostellar cluster (Sect. \ref{VLA1623}). The observations are taken as part of the ALMA Large Program (LP) Fifty AU STudy of the chemistry in the disk/envelope system of
Solar-like protostars (FAUST, http://faust-alma.riken.jp, PI: S. Yamamoto; \citealt{Codella2021}) (Sect. \ref{observations}). We analyse the spatial distribution of H$_2$CO,
HDCO, and D$_2$CO from scales of $\sim 2000$ au down to scales of 50 au (Sect. \ref{results}). We then estimate the gas temperature and the molecular column densities and deuteration towards the detected emission components, i.e. the VLA1623 A hot-corino, the outflow cavities, and the accretion streamers (Sect. \ref{D2}). Finally we discuss the formation mechanism of deuterated molecules and compare our estimates of deuteration in the different components of VLA1623 with values from the literature (Sects. \ref{comparison} and \ref{D1}). We summarize our conclusions in Sect. \ref{conclusions}.

\section{The VLA1623-2417 protostellar cluster}
\label{VLA1623}
VLA1623-2417 (hereafter VLA1623) is a protostellar cluster located in Ophiuchus A at a distance of 131$\pm$1 pc \citep{Gagne2018}. The multiple system
has been studied at different wavelengths
\citep[e.g.][and references therein]{Andre1990,Looney2000,Ward2011,Murillo2013disk,Murillo2018L,Murillo2018,Harris2018,Hsieh2020,Ohashi2022,Codella2022,Mercimek2023},
revealing four protostars: VLA1623 A, which in turn is a Class 0 binary protostar (A1 and A2) with a separation of $\sim$30 au, and surrounded by a circumbinary disk (CBD); VLA1623 B, a Class 0 protostar at $\sim$130 au from VLA1623 A; VLA1623 W, a Class I protostar located at $\sim$1300 au west of the VLA1623 A1+A2+B system.   

Complex gas kinematics characterize VLA1623. 
The VLA1623 A binary system drives blue- and red-shifted outflows revealed in CO and H$_{2}$ emission \citep[e.g.,][]{Andre1990, Hsieh2020, Hara2021} along the NW-SE direction. A rotating cavity opened by the mass loss process has been imaged in CS, CCH, and H$^{13}$CO$^+$ by \citet{Ohashi2022}. \citet{Santangelo2015} detected a fast jet driven by VLA1623 B. \citet{Murillo2018} and \citet{Ohashi2022} mapped a cold rotating cloud enveloping VLA1623 A and B in several simple species (e.g. DCO$^{+}$ and H$^{13}$CO$^{+}$). Recent ALMA studies imaged also molecular streamers infalling towards the A+B system in SO and C$^{18}$O lines \citep{Hsieh2020, Mercimek2023}. 
In the context of FAUST, \citet{Codella2022} mapped complex organic molecules, such as methanol (CH$_{3}$OH) and methyl formate (HCOOCH$_{3}$), which probe the circumstellar rotating gas around VLA1623 A1 and B. 
The detected rotation signatures suggest that the VLA1623 system is unstable: VLA1623 B counter-rotates with respect to the disk/envelope of VLA1623 A \citep[e.g.,][]{Ohashi2022, Codella2022}. 
Moreover, \citet{Murillo2013disk} and  \citet{Mercimek2023} discuss the possibility that the protostellar source VLA1623 W, which is located $\sim 1300$ au West to the A1+A2+B system and shows a systemic velocity which differs by 2.2 km s$^{-1}$, has been ejected from the close triple system.


\begin{table*}
\centering
     \begin{tabular}{lclccccccc}
        \hline
Species & Transition $^{a}$& $\nu$ $^{a}$ &   $E_{\rm up}$ $^{a}$ & $S\mu^2$ $^{a}$ & rms  & ${\delta} V$  & Beam size & Beam PA & FoV$^{b}$\\ 
& & (MHz)&(K) & (D$^{2}$) &  (mJy beam$\rm ^{-1}$) & (km s$^{-1}$) & ($\arcsec \times \arcsec$) & ($\rm ^{\circ})$ & $(\arcsec$)  \\
\hline
p-H$_{2}$CO &  3$_{0,3}$ -- 2$_{0,2}$ & 218222.2 & 21 &16 &4.4 &0.2 & 0.55 $\times$ 0.45&--74 & 24\\
HDCO & 4$_{1,4}$ -- 3$_{1,3}$ & 246924.6 &38 &20 & 0.7&1.2 &0.50 $\times$ 0.47 & --77 & 21 \\
HDCO & 4$_{2,2}$ -- 3$_{2,1}$ & 259034.9 & 63 & 16 &2.0 &0.2 &0.50 $\times$ 0.47  & --95 & 20\\
o-D$_{2}$CO &  4$_{0,4}$ -- 3$_{0,3}$& 231410.2 & 28 &43 &2.3 &0.2 &0.50 $\times$ 0.43 &--72 & 23\\
o-D$_{2}$CO &  4$_{2,3}$ -- 3$_{2,2}$ & 233650.4 & 50 &33 & 1.0 & 0.6 & 0.51 $\times$ 0.42 & --75 & 23 \\
        \hline
      \end{tabular}
       \caption{List of the observed lines towards VLA1623--2417. $^{a}$ Spectroscopic parameters are taken from the CDMS catalog \citep{Muller2005}. $^{b}$ The FWHM of Field of View. However the ALMA images cover larger area than FoV. }
\label{tab:lines}
    \end{table*}

\section{Observations}
\label{observations}

The VLA1623 protostellar system was observed between December 2018 and March 2020 by ALMA using Band 6 observations (216–234 GHz) in the context of the FAUST Large program (2018.1.01205.L; PI: Satoshi Yamamoto). Observations were conducted using two setups in Band 6 and one in Band 3. We used the 7-m array of the Atacama Compact Array (ACA) and the 12-m C43-4 and C43-1 configurations centering the observations at $\alpha_{\rm 2000}$ = 16${^h}$26${^m}$26${^s}$.392, $\delta_{\rm 2000}$ = --24$^\circ$24$'$30$''$.69. 
In this paper, we analyse only the observations in Band 6, i.e. in Setup 1 (214.0–219.0 GHz and 229.0–234.0 GHz) and Setup 2 (242.5–247.5 GHz and 257.2–262.5 GHz). 
Table \ref{tab:lines} reports the spectral parameters (transition, frequency, $\nu$, upper-level energy $E_{\rm up}$, and line strengths, $S\mu^2$) of the lines analysed in this paper: p-H$_2$CO (3$_{0,3}$ -- 2$_{0,2}$) (hereafter H$_2$CO), HDCO (4$_{1,4}$ -- 3$_{1,3}$), HDCO (4$_{2,2}$ -- 3$_{2,1}$), o-D$_{2}$CO (4$_{0,4}$ -- 3$_{0,3}$), and  o-D$_{2}$CO (4$_{2,3}$ -- 3$_{2,2}$) (hereafter D$_2$CO).
The molecular lines were observed with a bandwidth of 59 MHz (76-80 km s$^{-1}$) and a resolution of 122 kHz ($\sim$ 0.17 km s$^{-1}$), with the exception of the HDCO (4$_{1,4}$ -- 3$_{1,3}$) and o-D$_{2}$CO (4$_{2,3}$ -- 3$_{2,2}$) lines which were observed at lower spectral resolution in the broad spectral windows for the continuum (bandwidth of 1.9 GHz, and resolution of $1.2$ and $0.6$ km s$^{-1}$, respectively).

For each configuration, line-free continuum emission was used for self-calibration. We subtracted the continuum model derived from the self-calibration to produce continuum-subtracted line data. 
The quasars J1427-4206, J1517-2422, J1625-2527, J1924-2914, and J1626-2951 were used to calibrate the data, with a final absolute flux calibration uncertainty of $\sim$10\%.  We used a modified version of the calibration pipeline\footnote{https://github.com/autocorr/faust$\_$line$\_$imaging; Chandler et al. (in preparation)} within \textsc{CASA 5.6.1-8} \citep{CASA2022}, including an additional calibration routine to correct for $T_{\rm sys}$ issues and spectral data normalization\footnote{https://help.almascience.org/kb/articles/what-errors-could-originate-from-the-correlator-spectral-normalization-and-tsys-
calibration; Moellenbrock et al. (in preparation)}.

After merging the 12m and 7m configurations, continuum subtracted line cubes were cleaned using the task $tclean$. We adopted Briggs robustness parameter of 0.5 for molecular lines, while the parameter is -2.0 for the bright continuum to obtain the highest angular resolution. Primary beam corrections were also applied. 
The size and position angle (PA) of the synthesized beam, and the root mean square noise (r.m.s.) of the continuum maps in Setup 1 and Setup 2 are respectively: 0$\farcs$42 $\times$ 0$\farcs$32, PA=$-65^\circ$, 0.26  mJy\,beam$^{-1}$ and  0$\farcs$38 $\times$ 0$\farcs$34, PA=$67^\circ$, 0.27 mJy\,beam$^{-1}$. 
The details of the continuum-subtracted line datacubes are listed in Table \ref{tab:lines}: the r.m.s., the spectral resolution (${\delta} V$), the synthesized beam size and PA, and the field of view (FoV). 
  
\section{Results}
\label{results}

We detected emission from all the transitions of H$_2$CO and its isotopologues listed in  Table \ref{tab:lines}.
Figure \ref{spectra} shows the spectra of the detected transitions (in brightness temperature scale, $T_{\rm B}$) obtained by integrating the emission over the FoV. 

Figure \ref{Mom0MAPS} shows
moment 0 maps, i.e., the spatial distribution of the lines emission integrated over the emitting velocity range. The continuum emission at 1.3~mm is shown by magenta contours to pinpoint the positions of the A1 and A2 sources (not disentangled at the FAUST spatial resolution), their circumbinary disk, and the B and W protostars \citep[see also the FAUST continuum maps reported by][]{Codella2022, Codella2024, Mercimek2023}.
In the following subsections, we analyse the H$_{2}$CO, HDCO, and D$_{2}$CO emission and identify the structures probed by these species.

\subsection{H$_{2}$CO emission}

\begin{figure*}
\centering
\includegraphics[width=16cm]{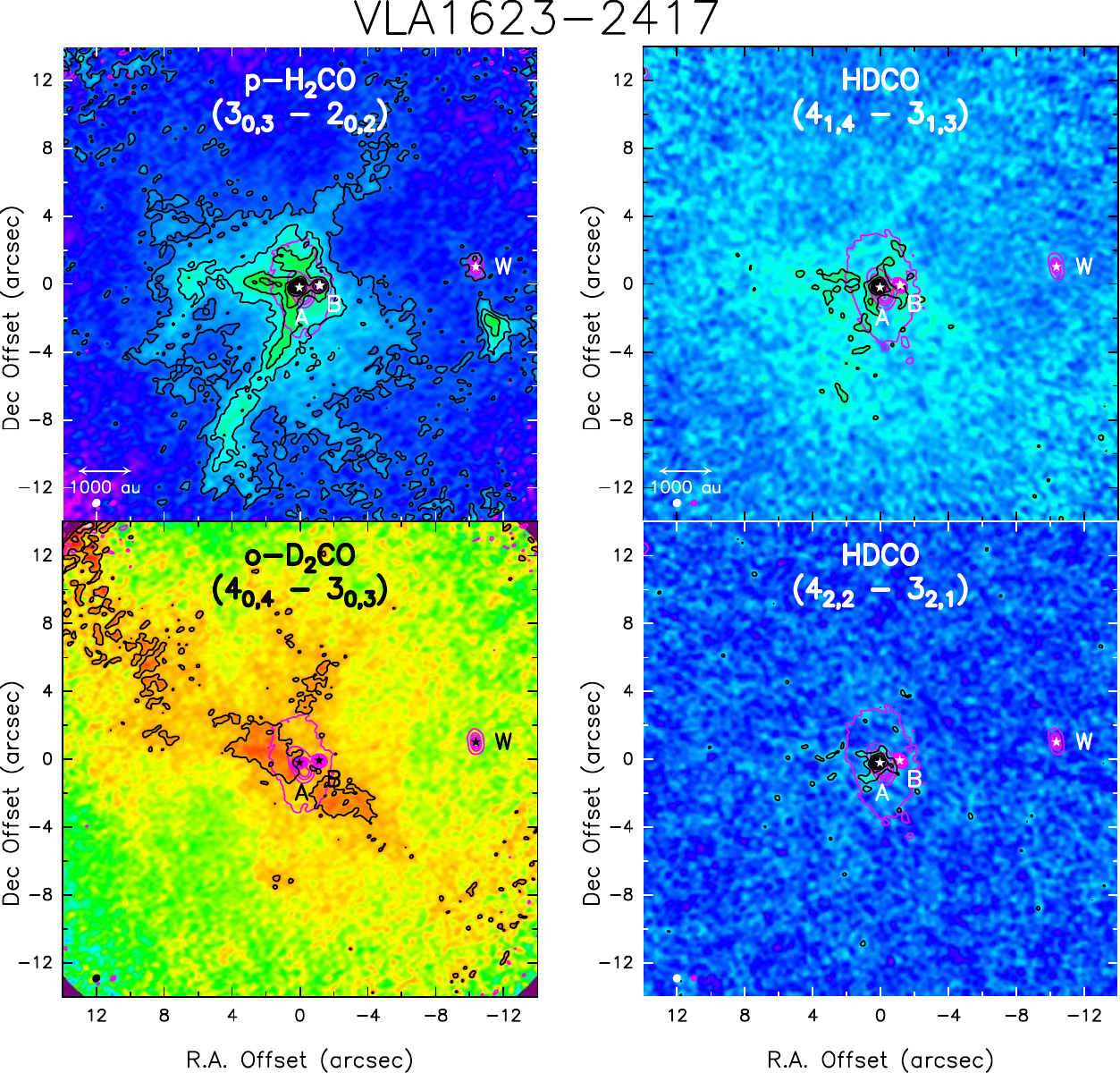}
\caption{Integrated intensity (moment 0) maps of H$_2$CO (3$_{0,3}$ -- 2$_{0,2}$), HDCO (4$_{1,4}$ -- 3$_{1,3}$), HDCO (4$_{2,2}$ -- 3$_{2,1}$), and D$_2$CO (4$_{0,4}$ -- 3$_{0,3}$) towards VLA 1623--2417. The maps have been obtained by integrating in the velocity range of [-5.2, +10.0], [+0.0, +6.2], [+0.0, +6.2], and [+3.0, +5.0] km s$^{-1}$, respectively. The white and black stars indicate the A, B, and W protostars. The first black contours and steps correspond to 3$\sigma$: 24.0, 15.0, 9.0, and 15.0 mJy km s$^{-1}$ beam$^{-1}$ for H$_{2}$CO (3$_{0,3}$ -- 2$_{0,2}$), HDCO (4$_{1,4}$ -- 3$_{1,3}$), HDCO (4$_{2,2}$ -- 3$_{2,1}$), and D$_{2}$CO (4$_{0,4}$ -- 3$_{0,3}$), respectively. The magenta contours are for the dust continuum emission
at 1.3~mm. For the continuum in Setup 1, which is shown in the H$_2$CO and D$_2$CO maps, the first contour and steps are 3$\sigma$ (0.78 mJy beam$^{-1}$) and 20$\sigma$, respectively. For the continuum in Setup 2, shown in the two HDCO maps, the first contour is 3$\sigma$ (0.81 mJy beam$^{-1}$) with steps of 20$\sigma$. The ellipses in the left bottom corners show the synthesized beams: in white and black for line emission (see Table \ref{tab:lines}) and in magenta for continuum emission.}
\label{Mom0MAPS}
\end{figure*}

Figure \ref{Mom0MAPS} (Upper-Left) shows the moment 0 map of H$_{2}$CO (3$_{0,3}$ -- 2$_{0,2}$) emission. We detect compact H$_2$CO emission towards all the protostars in the field, i.e. VLA1623 A, B, and W. In addition, formaldehyde probes
extended structures: the envelope of the  VLA1623 A+B system, the cavities opened by the outflow driven by VLA1623 A and  first imaged in CS $5-4$ by \citet{Ohashi2022}, and a bright elongated structure south of VLA1623 W.
In the following, we discuss the properties of both compact and extended emissions by focusing on the channel maps of H$_{2}$CO emission on both small scales (Sect. \ref{sec:smallscale}), and large scales (\ref{sec:largescale}).


\subsubsection{Compact H$_{2}$CO 
emission towards VLA1623 A, B, and W}
\label{sec:smallscale}

\begin{figure*}
\includegraphics[width=16cm]{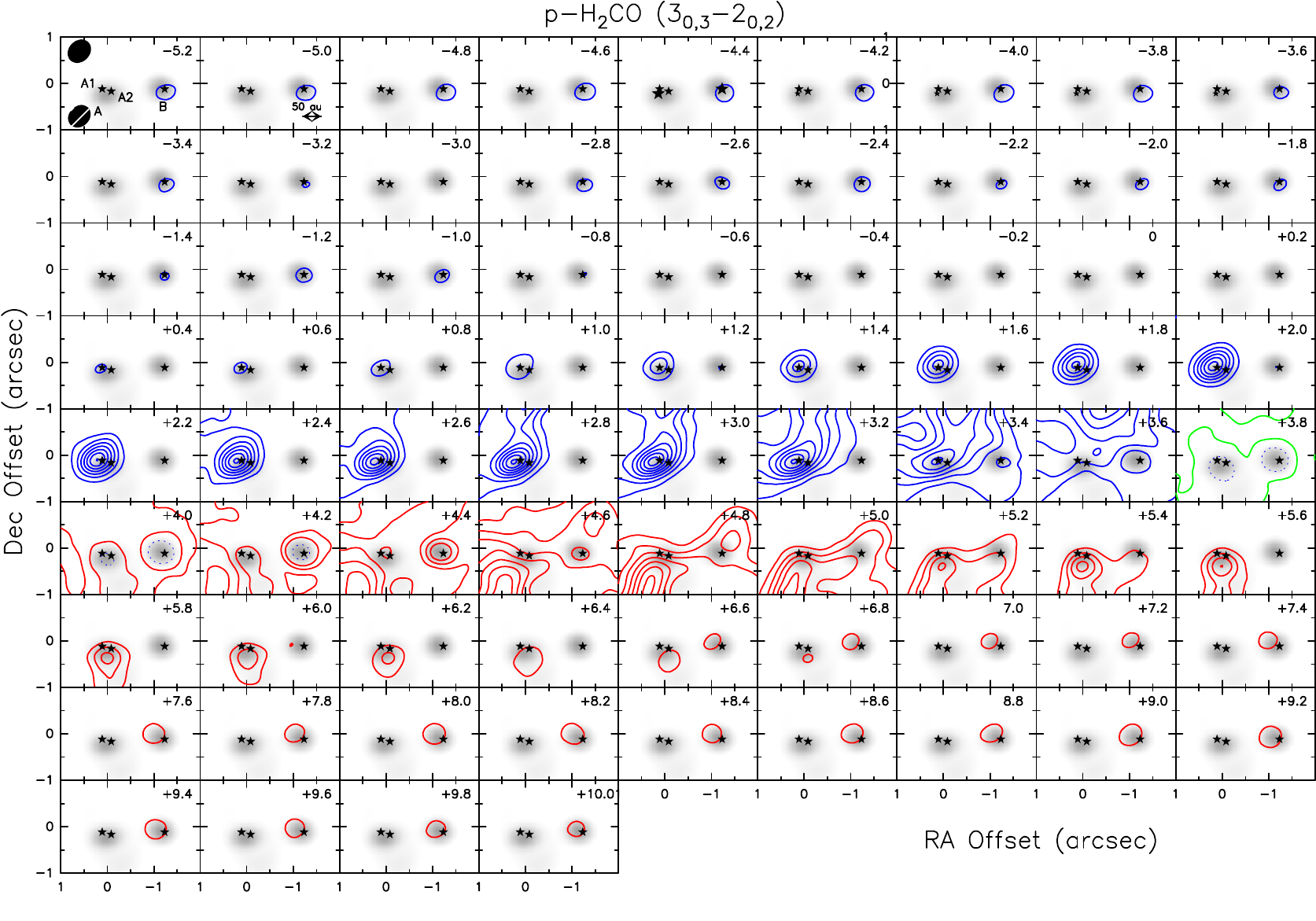}
\caption{Channel maps of the H$_2$CO (3$_{0,3}$ -- 2$_{0,2}$) emission in the $\sim$ 300 au region around the VLA1623 A1+A2 and B protostars (marked by the black stars). 
The LSR velocity  is labeled in the top-right corner of each channel. The contours of the emission are green in the channel at systemic velocity and blue/red in the channels at blue-/red-shifted velocities with respect to the systemic velocity of VLA1623 A and B \citep[+3.8 km s$^{-1}$][]{Ohashi2022}. 
The first contours and steps are 3$\sigma$ (13.2 mJy km s$^{-1}$ beam$^{-1}$). The grayscale background shows the 1.3~mm continuum emission in Setup 1. The dashed blue contour in the channels between +3.8 and +4.2 km s$^{-1}$ shows negative line intensity (absorption). The synthesized beams are shown in the top-left channel by filled and dashed ellipses for line and continuum emission, respectively.}
\label{H2CO_disk_AB}
\end{figure*}

The inspection of the channel maps of H$_{2}$CO allows us to reveal structures emitting only in selected velocity ranges, otherwise diluted in the moment 0 map. Figure \ref{H2CO_disk_AB} shows a zoom-in of the H$_{2}$CO (3$_{0,3}$ -- 2$_{0,2}$) emission on the $\sim$ 300 au region around the VLA1623 A and B sources. 
The emission towards the A and B protostars is compact and spatially unresolved ($\leq$ 50 au) at high-velocity, i.e. at blue- and red-shifted velocities larger than 2 km s$^{-1}$ with respect to the systemic velocity of the A and B sources, \citep[$V_{\rm sys}= +3.8$ km s$^{-1}$,][]{Ohashi2022} and shows velocity gradients, suggestive of rotation.
In particular, VLA1623 A shows red-shifted compact emission SW to the continuum peak (at [$+5.8$, $+6.8$] km s$^{-1}$, i.e. up to $+3$  km s$^{-1}$ with respect to $V_{\rm sys}$), and blue-shifted emission on the opposite side  ([$+0.4$, $+1.8$]  km s$^{-1}$, i.e. up to $-4.2$  km s$^{-1}$ with respect to $V_{\rm sys}$). The velocity gradient towards VLA1623 B is roughly along the same direction but is inverted (redshifted on the NE, and blue-shifted on the SW), and reach higher radial velocities, up to +10.0 km s$^{-1}$, and --5.2 km s$^{-1}$ (i.e. up to $+6.2$ km s$^{-1}$ and $-9$  km s$^{-1}$ with respect to $V_{\rm sys}$). 
Remarkably, the velocity gradients observed towards VLA1623 A and VLA1623 B are along the PA of their disks ($\sim 48-49^\circ$ for VLA1623 A1 and A2, and $\sim 42^\circ$ for VLA1623 B, from the high-resolution continuum maps by \citealt{Harris2018}), and are consistent with those shown in CS $5-4$ \citep{Ohashi2022}.
\citet{Codella2022} observed the same velocity gradients in CH$_3$OH lines towards A and B, suggesting that methanol probes the hot corinos (where dust mantles are sublimated at temperatures $\geq$ 100 K) or the outer regions of the protostellar
disks, where infalling gas from the envelope causes low-velocity shocks sputtering the dust \citep[][]{Sakai2014a, Sakai2014b}. 
However, the maximum radial velocities of H$_2$CO emission are smaller than those observed using CH$_3$OH
(up to $+7$ km s$^{-1}$ and $-2$ km s$^{-1}$ for VLA1623 A, and $+14$ km s$^{-1}$ and $-10$ km s$^{-1}$ for VLA1623 B), suggesting that formaldehyde emission originates from a more extended portion of the envelopes/disks associated with the A and B sources, with respect to methanol.
To conclude, the high-velocity H$_2$CO emission traces the counter-rotating envelopes around the A1+A2 binary system, and VLA1623 B.  
Note that Fig. \ref{H2CO_disk_AB} shows absorption towards the continuum peak of VLA1623 A and B, at velocities close to $V_{\rm sys}$, plausibly due to line absorption along the line of sight and optically thick dust (see Sect. \ref{D2}). 

\begin{figure*}
\includegraphics[width=16cm]{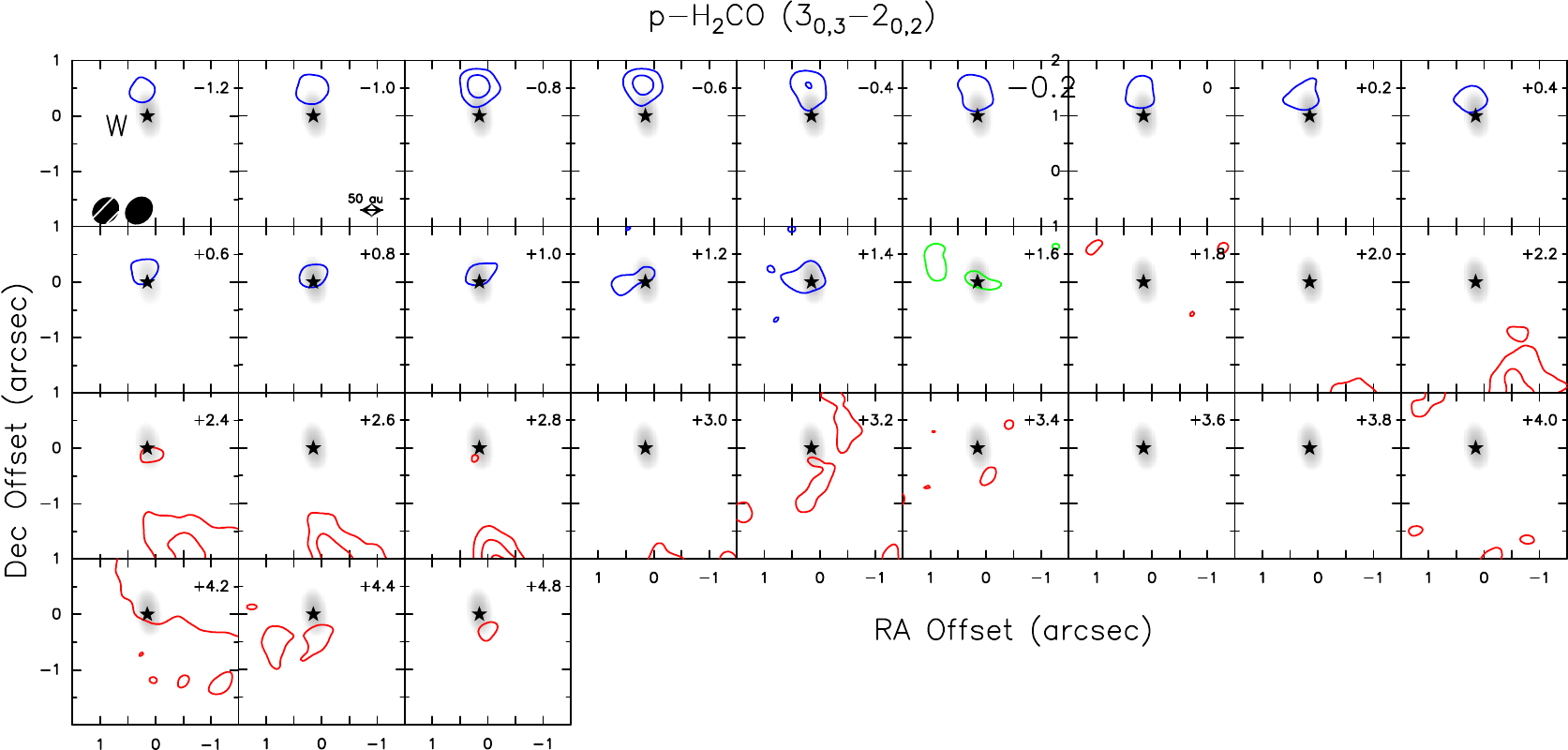}
\caption{Channel maps of the H$_2$CO (3$_{0,3}$ -- 2$_{0,2}$) emission in the $\sim$ 300 au region around the VLA1623 W protostar (marked by the black star).
The LSR velocity  is labeled in the top-right corner of each channel. The contours of the emission are green in the channel at systemic velocity and blue/red in the channels at blue-/red-shifted velocities with respect to the systemic velocity of VLA1623 W \citep[+1.6 km s$^{-1}$][]{Mercimek2023}.  
The first contours and steps are 3$\sigma$
(13.2 mJy km s$^{-1}$ beam$^{-1}$). The grayscale background shows the 1.3~mm continuum emission in Setup 1. The synthesized beams are shown in
the top-left channel by filled and dashed circles for line and continuum emission, respectively.}
\label{disk_sourceW}
\end{figure*}

Finally, Fig. \ref{disk_sourceW} shows the H$_{2}$CO (3$_{0,3}$ -- 2$_{0,2}$) channel maps in the $\sim$ 300 au region centered at the position of the VLA1623 W protostar.
The maps show spatially unresolved ($\leq$ 50 au) blue-shifted emission north of source W, up to velocities of --3 km s$^{-1}$ with respect to the systemic velocity of
VLA1623 W \citep[$V_{\rm sys}(W)=+1.6$ km s$^{-1}$,][]{Mercimek2023}.
This emission is consistent with the blue side of the inner envelope rotating along the N-S direction revealed by \citet{Mercimek2023} using the C$^{18}$O 2-1 line.
The non-detection of the southern red-shifted side in H$_2$CO is consistent with the C$^{18}$O emission, which is brighter in the blue-shifted lobe than in the red-shifted one. 

The compact H$_2$CO emissions towards the three protostars VLA1623 A, B, and W (labeled as A, B, and W), and the velocity ranges of emission are summarized in Table \ref{flows}.


\subsubsection{Streamers and Outflow Cavities in H$_{2}$CO emission}
\label{sec:largescale}

\begin{figure*}
\centering
\includegraphics[width=14cm]
{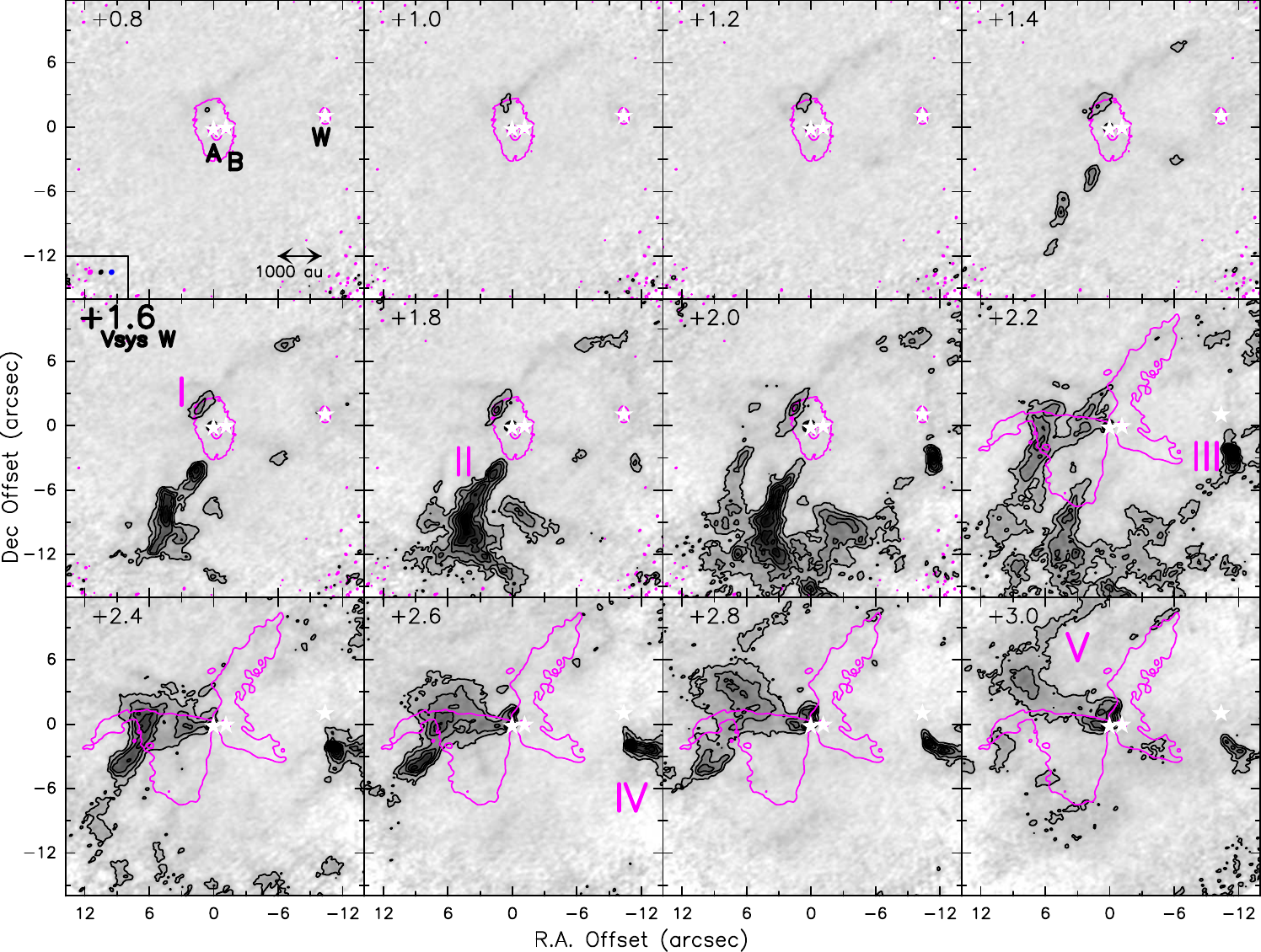}
\includegraphics[width=14cm]{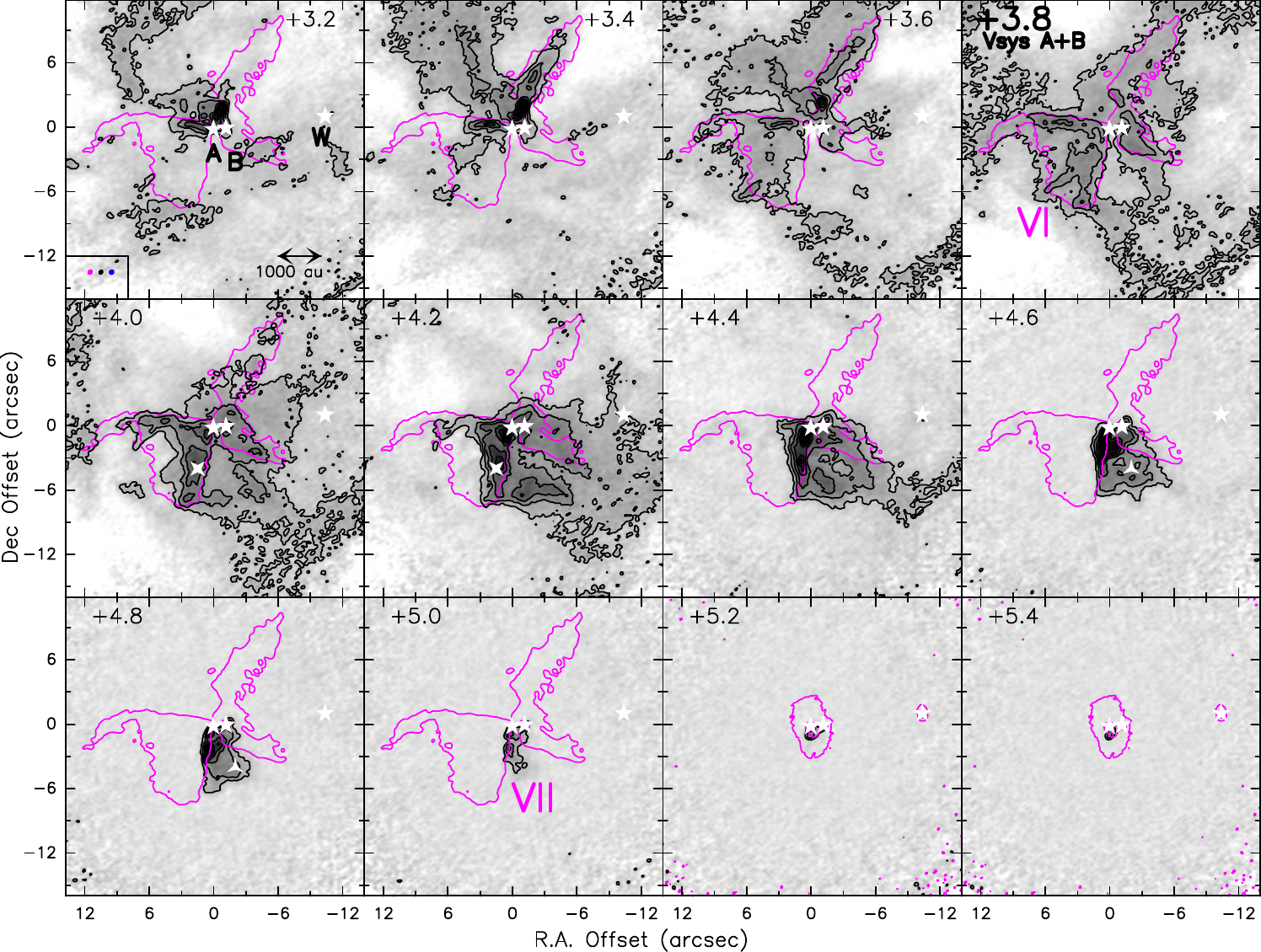}
    \caption{Channel maps of the H$_{2}$CO (3$_{0,3}$ -- 2$_{0,2}$) emission in the velocity range [+0.8, +5.4] km s$^{-1}$. The first contour is at 3$\sigma$, and the step is 3$\sigma$. The magenta contours in the channels from +2.2 km s$^{-1}$ to +5.0 km s$^{-1}$ indicate the outflow cavity walls probed by CS (5--4) emission  \citep[25$\sigma$ contour, from][]{Ohashi2022}. The magenta contour in the channels from +0.8 km s$^{-1}$ to +2.0 km s$^{-1}$ and from +5.2 km s$^{-1}$ to +5.4 km s$^{-1}$ indicate the dust continuum emission at 1.3~mm, starting from 3$\sigma$ (0.78 mJy beam$^{-1}$) with intervals of 40$\sigma$. The synthesized beams are shown by the magenta, black, and blue ellipses in the bottom-left corner of the first channel for H$_{2}$CO,  continuum, and CS (5--4) emission, respectively.
     The positions of VLA1623 A, B, and W are indicated by the white stars and are labeled in the first channel. The large-scale emission components listed as "I", "II", "III", "IV", "V", "VI", and "VII" in Table \ref{flows} are labeled. Cross and triangle symbols in the channels at $+4.0$ km s$^{-1}$, $+4.2$ km s$^{-1}$, and $+4.6$ km s$^{-1}$ indicate the positions along  the outflow cavity wall and the streamer where the line spectra are extracted (see Fig. \ref{superspectra} in Sect. \ref{D2}). }
    \label{channel_large_H2CO}
\end{figure*}

Figure \ref{channel_large_H2CO} shows the H$_{2}$CO (3$_{0,3}$ -- 2$_{0,2}$) velocity channel maps over a larger region ($\sim  24\arcsec \times 24\arcsec$) with respect to what reported in Sect. \ref{sec:smallscale}.
As references, we draw in magenta the outflow cavity walls probed by CS(5–4) emission \citep{Ohashi2022}, and the dust continuum emission at 1.3~mm.
Several extended structures are detected in H$_{2}$CO.
Table \ref{flows} lists these structures, and the velocity ranges over which they are detected in the channel maps, moving from blue-shifted to red-shifted velocities, and labeling them from I to VII: 

\begin{itemize}

\item \textbf{(I)} in the [+0.8, +2.2] km s$^{-1}$ velocity channels, we detect H$_2$CO emission at the northern edge of the circumbinary disk surrounding VLA1623 A1+A2. This component
has already been detected in SO lines by
\citet{Hsieh2020} and \citet{Codella2024}. They suggested that
this feature is associated with an accretion shock produced by a blue-shifted streamer infalling onto the circumbinary disk (component V, see below);

\item \textbf{(II)} the emission on the [+1.4 and +2.2] km s$^{-1}$ velocity range reveals an extended structure located south to VLA1623 A. This component, already imaged in SO lines by \citet{Codella2024}, is a blue-shifted streamer infalling on the VLA1623 B protostar. Although the position of the streamer (projected on the plane of the sky) could suggest its association with 
the southern outflow cavity of VLA1623 A, this is excluded by the fact that the accretion streamer is blue-shifted while the southern cavity side is red-shifted;

\item \textbf{(III)} a streamer connecting the southern region around A+B with the VLA1623 W protostar is detected in the [+1.8, +3.2] km s$^{-1}$ velocity range. The streamer is red-shifted with respect
to the systemic velocity of the W object \citep[+1.6 km s$^{-1}$:][]{Mercimek2023}. 
This streamer has been previously detected in C$^{18}$O by \citet{Mercimek2023}. The H$_{2}$CO emission traces the streamer without the contamination of the A+B envelope emitting in C$^{18}$O; 

\item \textbf{(IV)} the [+2.4, +3.2] km s$^{-1}$ velocity range shows an additional elongated structure SW of VLA1623 W. This structure is detected close to the half-power point of the primary beam, therefore no reliable conclusion on its origin can be drawn; 

\item \textbf{(V)} a slightly blue-shifted streamer extending from the North edge of the FoV to NW and then pointing towards VLA1623 A is imaged in the [+2.8, + 3.6] km s$^{-1}$ velocity range. This emission likely traces a streamer infalling onto the circumbinary system and causing an accretion shock, whose emission is labeled as I. This scenario has already been 
suggested by \citet{Hsieh2020} based on SO and C$^{18}$O lines.

\item \textbf{(VI)} at velocities close to the systemic velocity of VLA1623 A and B, i.e. in the [+3.4, +4.4] km s$^{-1}$ velocity range, H$_2$CO emission probes the region encompassed by the rotating outflow cavities first imaged in CS $5-4$ by \citet{Ohashi2022}. The association with the outflow cavity is confirmed by kinematics: the northern side is blue-shifted, while the southern side is red-shifted, in agreement with CS emission.
The H$_2$CO emission as a probe of outflow cavities has been reported also by \citet{sabatini24} for the outflow associated with the IRS7B protostellar system in the Corona Australis. 

\item \textbf{(VII)} a red-shifted elongated structure south of VLA1623 A is detected at [+4.4, +5.0] km s$^{-1}$ velocities, which is not coincident with the southern outflow cavities. 
Thie emission component is perpendicular to the outflow direction, and is plausibly associated with a streamer infalling onto the VLA1623 A and B protostars. \citet{Hsieh2020} has previously reported this structure, proposing the occurrence of an accretion streamer feeding the (red-shifted) southern side
of the circumbinary disk around VLA1623 A.

\end{itemize}

\begin{table*}
\centering
     \begin{tabular}{lccccc}
     \hline
Label & Component & H$_2$CO & HDCO & D$_2$CO & Spectrum Position \\
      &           & (km s$^{-1}$) & (km s$^{-1}$) & (km s$^{-1}$) & (\arcsec, \arcsec)\\
\hline
\multicolumn{6}{c}{COMPACT} \\
A   & inner rotating envelope of A & [$+0.4$, $+1.8$]  & [$+0.8$, $+2.6$] & -- & (0, 0)\\
    &                              & [$+5.8$, $+6.8$]  & [$+5.0$, $+6.8$] & -- & \\
B   & inner rotating envelope of B & [$-5.2$, $-1.0$]  & --                & -- & ($-1.16$, $+0.11$) \\
    &                              & [$+6.6$, $+10.0$] & --                & -- & \\
W   & inner rotating envelope of W & [$-1.2$, $+1.6$]  & --                & -- & ($-10.32$, $+1.26$)  \\
    &                              & --                 & --                & -- & \\
\hline
\multicolumn{6}{c}{EXTENDED} \\
I   & accretion shock at the northern edge of A CBD & [$+0.8$, $+2.2$] & -- & -- & -- \\
II  & southern blue-shifted streamer towards B      & [$+1.4$, $+2.2$] & -- & -- & -- \\
III & southern red-shifted streamer towards W       & [$+1.8$, $+3.2$] & -- & -- & -- \\
IV  & SE elongated emission associated with W       & [$+2.4$, $+3.2$] & -- & -- & -- \\
V   & northern blue-shifted streamer towards A      & [$+2.8$, $+3.6$] & [$+2.6$, $+3.8$] & [$+3.2$, $+3.6$] & -- \\
VI  & outflow cavities driven by A                  & [$+3.4$, $+4.4$] & [$+2.6$, $+5.0$] & [$+3.4$, $+4.4$] & ($+1.49$,  $-3.78$)\\
VII & southern red-shifted streamer towards A       & [$+4.4$, $+5.0$] & [$+3.8$, $+5.0$] & [$+4.4$, $+5.0$] & ($-2.01$, $-3.78$) \\
        \hline
     \end{tabular}
\caption{Compact and extended emission components towards VLA1623-2417 identified in H$_2$CO, HDCO, and D$_2$CO channel maps in the listed velocity ranges. The positions (R.A. and Dec. offsets with respect to VLA1623 A) where the spectra of the identified components are extracted are listed in the last column.}
\label{flows}
\end{table*}


\subsection{HDCO emission}

The moment 0 maps of the HDCO (4$_{1,4}$ -- 3$_{1,3}$) and (4$_{2,2}$ -- 3$_{2,1}$) lines are shown in Fig. \ref{Mom0MAPS}, while the channel maps over a small ($\sim 3\arcsec \times 2\arcsec$) and large ($\sim 19\arcsec \times 16\arcsec$) region are shown in Fig. \ref{HDCO_38K_APPENDIX} and \ref{HDCO_63K_APPENDIX}.
The maps show no HDCO emission detected toward VLA1623 B and W, while the emission towards VLA1623 A consists of several components.

In Figure \ref{2HDCO} we show the HDCO emission integrated over selected velocity intervals, i.e. at systemic velocity ($V_{\rm sys}$ = +3.8 km s$^{-1}$, left panels); at low-velocity (up to $\pm$ 1.2 km s$^{-1}$ with respect to $V_{\rm sys}$, middle panels);  and at high velocity  (up to $\pm3.0$ km s$^{-1}$ with respect to $V_{\rm sys}$, right panels). 
The figure shows that some of the components identified in the analysis of the H$_2$CO channel maps are also associated with HDCO emission, namely: 
\begin{itemize}
\item \textbf{(A)} the high-velocity HDCO emission is compact around the A1+A2 binary system, and show a velocity gradient along the NE-SW direction, in  agreement with H$_2$CO (see Fig. \ref{H2CO_disk_AB}), and CH$_3$OH \citep{Codella2022}. 
As for H$_2$CO, the HDCO compact emission is at  lower velocities than methanol, thus it probes a more extended region of the rotating envelope around VLA1623 A (see Sect. \ref{sec:smallscale});
\item \textbf{(VI)} the HDCO emission at systemic velocity and low blue- and red-shifted velocities traces the rotating outflow cavities driven by VLA1623 A, in agreement with CS and H$_2$CO emission \citep{Ohashi2022}; 
\item \textbf{(V, VII)} the lower excitation HDCO (4$_{1,4}$ -- 3$_{1,3}$) line ($E_{\rm u}$ = 38 K), which is brighter, also shows low-velocity emission north and south of VLA1623 A,  which is not associated with the outflow cavities.
Despite the low S/N ($\sim$ 3$\sigma$) of the HDCO channel maps, we identify that the low-velocity  emission is associated with the northern blue-shifted and southern red-shifted streamers infalling on VLA1623 A and revealed by H$_2$CO (components V and VII, see Table \ref{flows}).
\end{itemize}

\subsection{D$_2$CO emission}

The moment 0 map of the D$_2$CO lower excitation line (4$_{0,4}$ -- 3$_{0,3}$) is shown in Fig. \ref{Mom0MAPS}, while Figure \ref{channel_D2CO} shows the velocity channel maps, with overplot the outflow cavities revelead by CS emission \citep{Ohashi2022}, and the dust continuum emission at 1.3~mm. Table \ref{flows} lists the detected structures, moving along
the velocity channel maps from blue-shifted to red-shifted velocities, and labeling them from V to VII. 
As for HDCO, D$_{2}$CO shows no emission towards the VLA1623 B and W protostars (components B and W in Table \ref{flows}), while faint emission from the inner envelope of VLA1623 A is recovered in the spectrum extracted at the source continuum peak (see Sect. \ref{D2}).
Surprisingly, the D$_{2}$CO moment 0 map and channel maps reveal a spatial distribution which is different than that of HDCO and H$_2$CO: the D$_{2}$CO emission close to systemic velocity ($V_{\rm sys}$ $\pm$ 1 km s$^{-1}$) extends out to the edges of the field of view, and probe a region which does not coincide with  the wide-angle outflow cavities outlined by the CS emission. 
Figure \ref{flows_D2CO_H2CO} compares the spatial distributions of H$_{2}$CO and D$_{2}$CO in selected velocity ranges, and allows us to identify some of the structures detected in H$_2$CO and reported in Table \ref{flows}.
More specifically, D$_2$CO probes the following components: 
\begin{itemize}
    \item {\bf (V)} on the [+3.2, +3.6] km s$^{-1}$ velocity range (Panel a) the D$_2$CO emission is partially coincident with the northern blue-shifted streamer revealed by H$_2$CO. However, the peak of the D$_2$CO emission is at the edges of the H$_2$CO structure;  
    \item {\bf (VI)} the D$_2$CO emission on the [+3.4, +4.4] km s$^{-1}$ velocity range (Panel b) traces  the outflow cavities detected in H$_2$CO. Note that the D$_2$CO emission at low velocities looks more contaminated by extended emission plausibly associated with the large-scale envelope; 
    \item {\bf (VII)} the D$_2$CO emission on [+4.4, +5.0] km s$^{-1}$ velocity range (Panel c) appears associated with the southern red-shifted streamer labeled as VII in the H$_2$CO channel maps. Figure \ref{D2CO_Mom1and2} shows the intensity-weighted mean velocity (moment 1) map of D$_2$CO (4$_{0,4}$ -- 3$_{0,3}$) emission, where the red-shifted streamer infalling on the VLA1623 A system from the SE is clearly outlined.
    However, also in this case, the emission peaks of H$_2$CO and D$_2$CO are not colocated,
    with H$_2$CO mainly associated with the eastern side of the VII structure and D$_2$CO mainly emitted from the western side.   
\end{itemize}

The higher excitation D$_{2}$CO (4$_{2,3}$ -- 3$_{2,2}$) line ($E_{\rm up} \sim 50$ K) is also marginally detected. 
The channel maps (Fig. \ref{channel_D2CO_50K}) show emission at very low S/N ($\sim$ 2--3$\sigma$), but tentatively suggest
a spatial distribution consistent with that of the brighter lower-excitation D$_{2}$CO line (4$_{0,4}$ -- 3$_{0,3}$). 
The maps show signatures of the outflow cavities at systemic velocity ($+3.8$ km s$^{-1}$), and of the southern red-shifted streamer at $+4.4$ km s$^{-1}$ (labeled as VI and VII in Table \ref{flows}). 
Despite the low S/N, the D$_{2}$CO (4$_{2,3}$ -- 3$_{2,2}$) line is clearly detected in the spectrum obtained by integrating the emission over a  10$\arcsec$ region centered on VLA1623 A (see Fig. \ref{D2CO-233}). The line profile 
is in agreement with the profile of the lower excitation (4$_{0,4}$ -- 3$_{0,3}$) line (see Fig. \ref{superspectra}), peaking in the same velocity range (3.8 -- 4.4 km s$^{-1}$).
\\

\section{Discussion}
\label{discussion}

The analysis of the channel maps presented in Sect. \ref{results} showed that H$_2$CO and its deuterated isotopologues trace several components of the 
VLA1623 protostellar system: the inner protostellar envelopes, the  outflow cavities, and the accretion streamers.  
VLA1623 is therefore an excellent laboratory to study the D/H ratio in these different components.
The derivation of the D/H values is reported in Sect. \ref{D2}, while in Sect. \ref{comparison} we compare the estimated D/H ratios with values derived for other sources in the literature.

In addition, within a fixed emitting velocity range, H$_2$CO, HDCO, and D$_2$CO do not always show the same spatial distribution. The origin of this difference will be discussed in Sect. \ref{D1}. 

\begin{figure*}
\centering
\includegraphics[width=17cm]{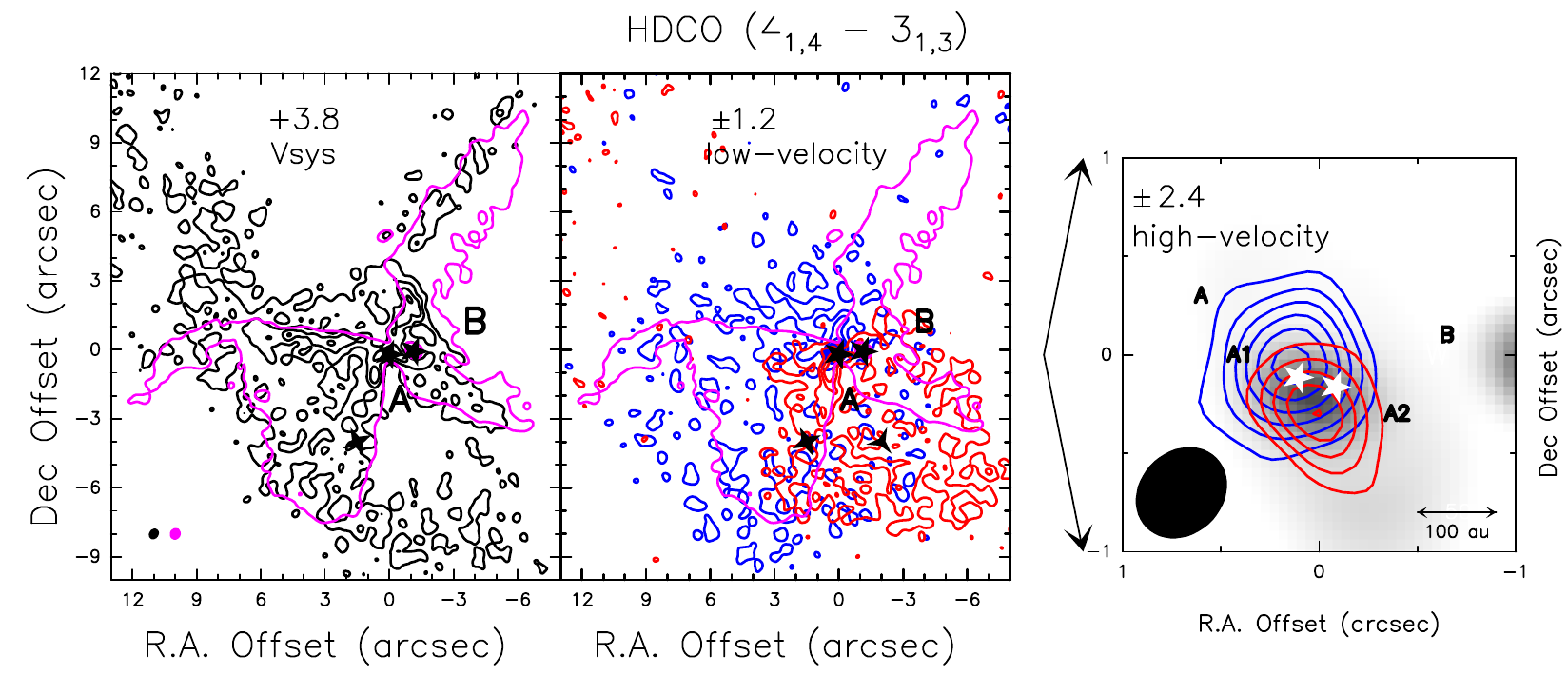}
\includegraphics[width=17cm]{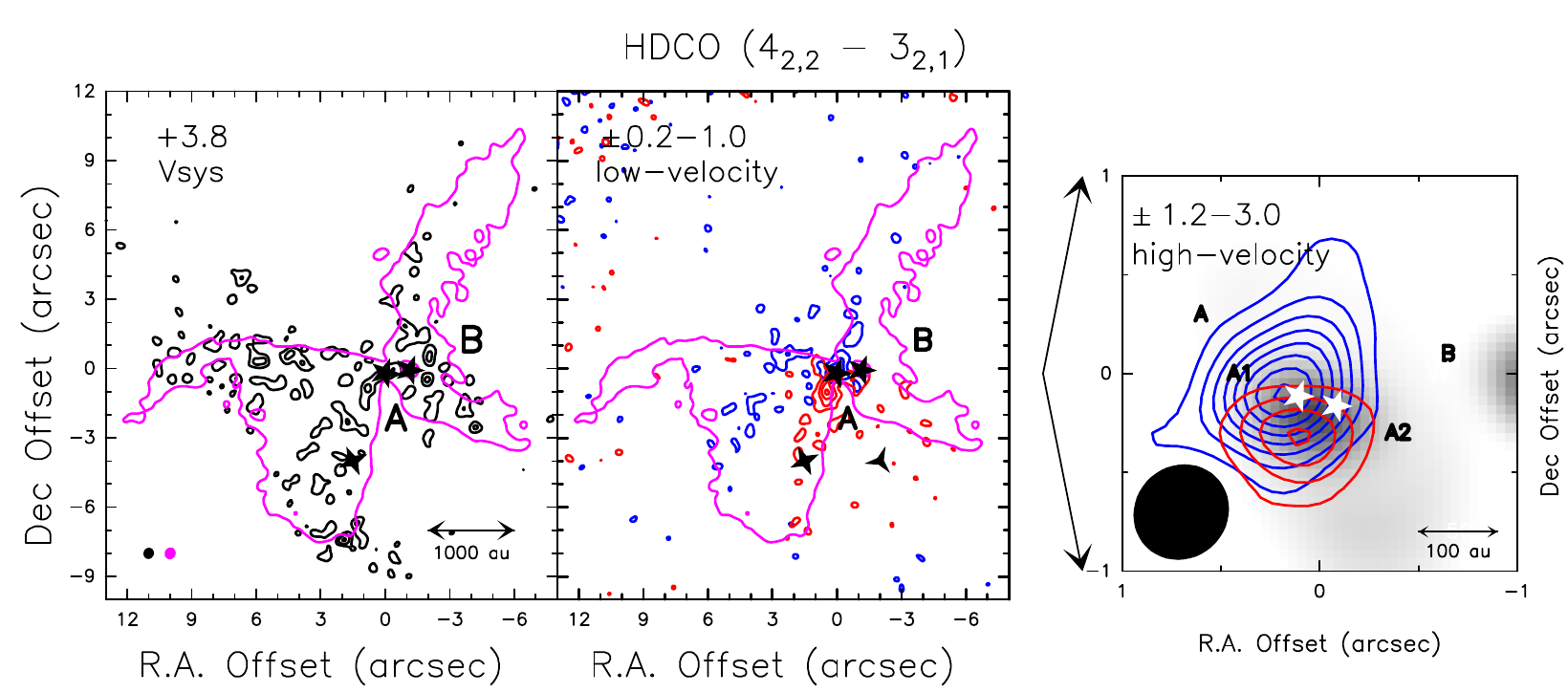}
    \caption{Velocity-integrated maps of HDCO lines towards VLA1623. \textit{Top row:}  HDCO (4$_{1,4}$ -- 3$_{1,3}$). The left panel shows the emission at systemic velocity ($+3.8$ km s$^{-1}$, in black contours), the middle and right panels show low-velocity ($\pm$1.2 km s$^{-1}$) and high-velocity ($\pm$2.4 km s$^{-1}$) emission (red and blue contours). The first contours and steps are 3$\sigma$ (2.1 mJy beam$^{-1}$). \textit{Bottom row:} HDCO (4$_{2,2}$ -- 3$_{2,1}$). The left panel shows the emission at systemic velocity  (+3.8 km s$^{-1}$, in black contours), the middle and right panels shows low-velocity ([$\pm$0.2, $\pm$1.0] km s$^{-1}$) and high-velocity ([$\pm1.2$, $\pm3.0$] km s$^{-1}$) emission (red and blue contours). The first contours and steps are 3$\sigma$ (3$\sigma$ is 6.0 and 6.0 mJy beam$^{-1}$ for low- and high-velocity maps, respectively.). The white stars in the right panels (high-velocity emission maps) indicate the position of VLA1623 A1, A2 from  \citet{Harris2018}; the black stars in the other panels show the position of VLA1623 A and B from the FAUST continuum maps at 1.3~mm  \citep{Mercimek2023}. The magenta contours show the CS (5--4) emission which probes the outflow cavity walls \citep[25$\sigma$ contour, from][]{Ohashi2022}. The grayscale background shows the 1.3 mm continuum emission in Setup 2. The HDCO and CS synthesized beams are shown in black and magenta in the bottom-left of the left and right panels. Cross and triangle symbols in the left and middle panels show the positions of the outflow cavity and streamer where the spectra are extracted (see Fig. \ref{superspectra} in Sect. \ref{D2}).
    }
   \label{2HDCO}
\end{figure*}

\begin{figure*}
\centering
\includegraphics[width=14cm]
{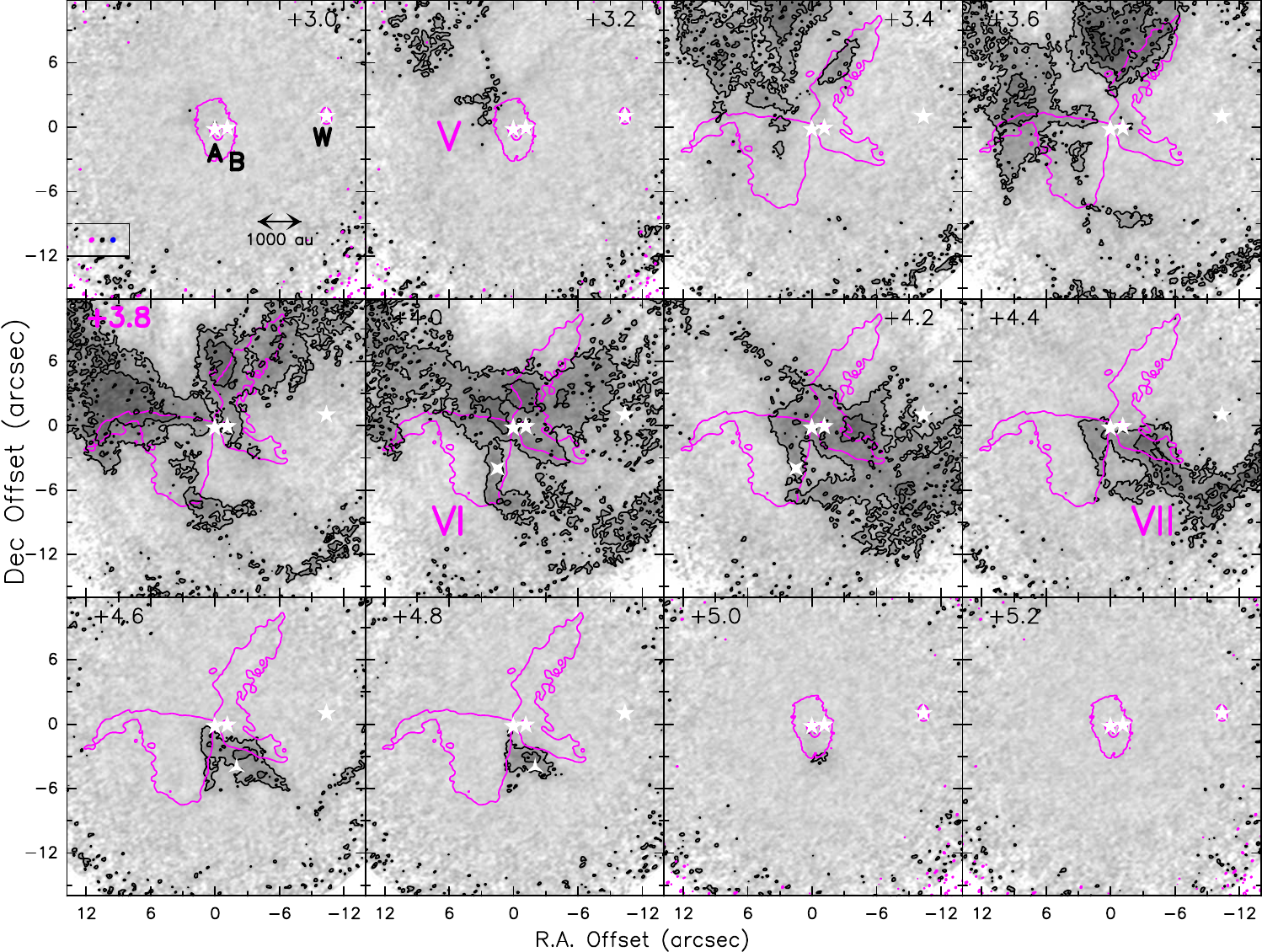}
    \caption{Channel maps of the D$_2$CO (4$_{0,4}$ -- 3$_{0,3}$) emission in the velocity range [+3.0, +5.2] km s$^{-1}$. The first contour is at 3$\sigma$, and the step is 3$\sigma$. The magenta contours in the channels from +3.4 km s$^{-1}$ to +4.8 km s$^{-1}$ indicate the outflow cavity walls probed by CS (5--4) emission  \citep[25$\sigma$ contour, from][]{Ohashi2022}. The magenta contour in the channels from +3.0 km s$^{-1}$ to +3.2 km s$^{-1}$ and from +5.0 km s$^{-1}$ to +5.2 km s$^{-1}$ indicate the dust continuum emission at 1.3~mm, starting from 3$\sigma$ (0.78 mJy beam$^{-1}$) with intervals of 40$\sigma$. The synthesized beams are shown by the magenta, black, and blue ellipses in the bottom-left corner of the first channel for CS (5--4), continuum, and H$_{2}$CO, respectively.
     The positions of VLA 1623A, B, and W are indicated by the white stars and are labeled in the first channel.
     The large-scale emission components are labeled as  "V," "VI," and "VII" in Table \ref{flows}, consistently with the H$_2$CO components. Cross and triangle symbols at channels between +4.2 and +4.8 km s$^{-1}$ show where the spectra of outflow cavity wall and streamer are extracted (see Fig. \ref{superspectra} in Sect. \ref{D2}).}
    \label{channel_D2CO}
\end{figure*}

\begin{figure}
\includegraphics[width=8.5cm]{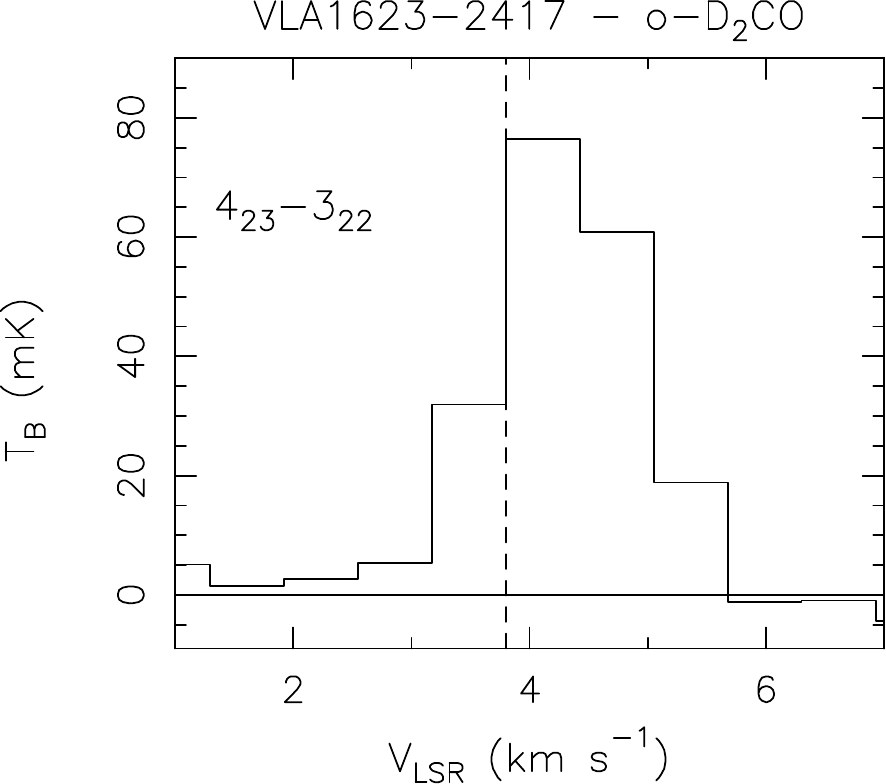}
\caption{Spectral profile of D$_2$CO (4$_{2,3}$ -- 3$_{2,2}$) profile
(in $T_{\rm B}$ scale), integrated on a $10\arcsec$ region centered on VLA1623 A. The dashed vertical line indicates the systemic velocity of VLA1623 A and B (+3.8 km s$^{-1}$, \citealt{Ohashi2022}). }
\label{D2CO-233}
\end{figure} 

\subsection{Deuteration in the inner envelope, in the outflow cavities, and in the streamers}
\label{D2}
\begin{figure*}
\centering
\includegraphics[width=19cm]
{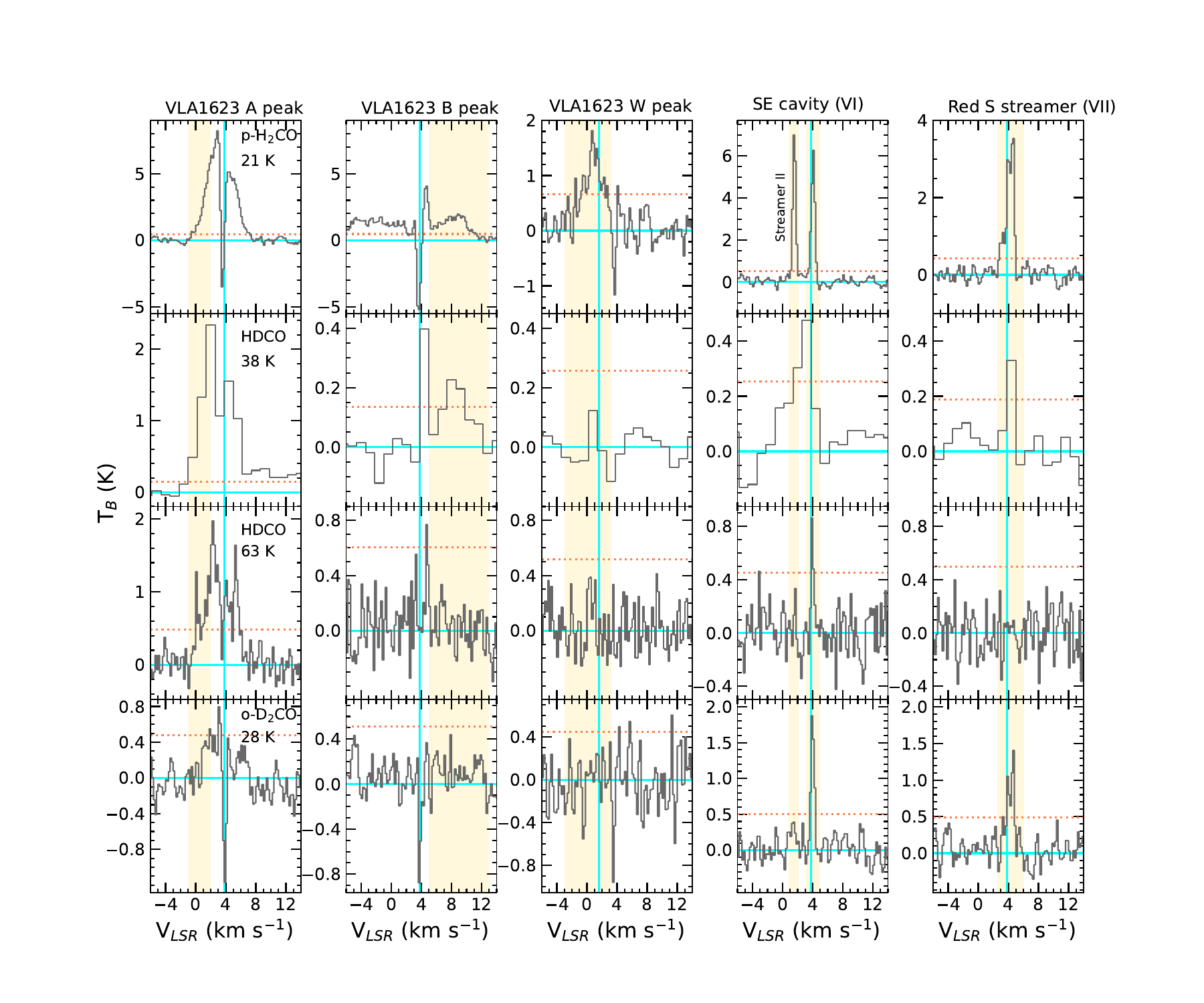}
    \caption{Spectra of the H$_2$CO, HDCO, and D$_{2}$CO lines extracted at several positions of the VLA1623-2417 cluster (in brightness temperature, $T_{\rm B}$). Because of the low S/N, for the low-excitation  D$_2$CO (4$_{2,3}$ -- 3$_{2,2}$) line we report only the spectrum integrated on a $10\arcsec$ region (see Fig. \ref{D2CO-233}).
    The positions are (from left to right): dust continuum peak of VLA1623 A, B, and W (first, second, and third columns, respectively); a position along the SE cavity (VI components, fourth column); a position along the red-shifted southern streamer (VII component, fifth column). The vertical solid line stands for the source systemic velocity ($+3.8$ km s$^{-1}$ for VLA1623 A and B, and $+1.6$ km s$^{-1}$ for VLA1623 W). The orange horizontal dotted line indicates the 3$\sigma$ level. The blue-shifted emission peak in the H$_2$CO spectrum of the SE outflow cavity (VI) traces the southern blue-shifted streamer, labeled II in Table \ref{flows}. Yellow bands show the velocity ranges of integration of the line intensities reported in Table \ref{temperatures}.
    }
    \label{superspectra}
\end{figure*}

Figure \ref{superspectra} shows the spectra (in $T_{\rm B}$ scale) of the observed H$_2$CO, HDCO, and D$_2$CO lines (see Tab. \ref{tab:lines}), extracted in five selected positions representative of the compact and extended components identified towards the VLA1623-2417 cluster:
at the dust continuum peaks of the protostellar sources VLA1623 A, B, and W (components A, B, and W in Table \ref{flows}); at a representative position of the SE outflow cavity (VI component); and at a representative position of the red-shifted southern streamer (VII component). 
The R.A. and Dec. offset of the selected positions are listed in Tab. \ref{flows}, and marked in Figs. \ref{channel_large_H2CO}, \ref{2HDCO}, \ref{channel_D2CO}.

Different positions are associated with different line profiles,
reflecting the kinematical components shown by the channel 
maps analysed in Sect. \ref{results}:

\begin{itemize}
\item[-] The spectra extracted at the position of the VLA1623 A continuum peak exhibit a narrow absorption feature at systemic velocity detected in the lower excitation H$_2$CO and D$_2$CO lines ($E_{\rm up}$ = 21 K, and 28 K). This is due to absorption by the cold gas along the line of sight and by the optically thick dust continuum emission.
The spectra shows blue-shifted and red-shifted emission, which probe the inner rotating envelope of A. In additon, the lines profiles are asymmetric with the blue-shifted emission more intense than the red-shifted one, suggestive of infall.

\item[-] The H$_2$CO spectrum toward VLA1623 B shows symmetric red-shifted and blue-shifted emission out to $\sim \pm 10$ km s$^{-1}$ with respect to the systemic velocity (+3.8 km s$^{-1}$), which probes the inner portion of the rotating envelope of B,  consistently with the H$_2$CO channel maps (Fig. \ref{H2CO_disk_AB}). 
As for the spectra towards VLA1623 A, the lower excitation H$_2$CO and D$_2$CO lines  exhibit a narrow absorption feature at systemic velocity due to absorption by the cold gas along the line of sight and by the optically thick dust continuum emission.
This absorption feature has been also observed in the CS, CCH, and H$^{13}$CO$^{+}$ 
low excitation lines ($E_{\rm up}$ = 25-35 K) toward VLA1623 B \citep{Ohashi2022}. 
In addition, the H$_2$CO spectrum shows a red-shifted excess, also observed in HDCO, which originates from the NW red-shifted side of the rotating outflow cavity associated with VLA1623 A (labeled as NW Outflow Cavity in Table \ref{temperatures}). 

\item[-] The H$_2$CO spectrum towards VLA1623 W shows blue-shifted emission, unveiling the blue-shifted northern side of the inner rotating envelope of W, as illustrated in Fig. \ref{H2CO_disk_AB}, which is brighter than the red-shifted one consistent with the observations of C$^{18}$O by \citet{Mercimek2023}. No emission is detected in HDCO and D$_2$CO.

\item[-] The H$_2$CO spectrum extracted at the position along the SE outflow cavity (component VI) peak at systemic velocity and shows a counterpart in the HDCO and D$_2$CO spectra. In addition, the spectrum shows a blue-shifted emission component, which probes the southern blue-shifted streamer (component II), which spatially overlaps with the cavity. This component is also detected in HDCO.

\item[-] The spectra extracted at the position of the southern red-shifted streamer (component VII) exhibits a narrow (0.6--0.8 km s$^{-1}$) nearly Gaussian emission component at red-shifted velocities in H$_2$CO, HDCO, and D$_2$CO. 
\end{itemize}

Despite the narrow range of upper-level energies of the detected HDCO lines (38 K and 63 K), we used the intensity ratio of the two HDCO lines to determine the excitation temperature ($T_{\rm ex}$) of the gas in the different components.
Local thermodynamic equilibrium (LTE) and optically thin emission have been assumed. To properly measure the line intensities from the spectra extracted at the position of the continuum peaks of A and B and avoid the effect of absorption at systemic velocity, we integrated the emission on the blue-shifted and red-shifted velocity ranges highligthed in Fig. \ref{superspectra} by yellow bands. Table \ref{temperatures} reports, for each component, the velocity ranges used for integration, the obtained line integrated intesities, and the  excitation temperature estimated from the HDCO line ratio. 

The excitation temperature towards VLA1623 A is $125\pm$60 K, supporting the association of the observed line emission with the inner envelope heated by the VLA1623 A binary system.
The excitation temperature in the 
SE outflow cavity is lower 37$\pm$19 K. Low $T_{\rm ex}$ is obtained also towards  VLA1623 B
(28$\pm$12 K), confirming that the HDCO emission detected towards B plausibly probes the redshifted side of the rotating outflow cavity as suggested by the maps in Fig. \ref{2HDCO}.
Finally, the excitation temperature estimated in the streamers (II and VII) is an upper limit, due to the non-detection of the higher excitation HDCO line, and is $\le15$ K. This is consistent with the expected temperatures at a separation of $\sim600$ au  from the protostars (see Tab. \ref{flows}).

To determine the column densities of H$_2$CO, HDCO, and D$_2$CO, we assumed LTE and optically thin emission and adopted the excitation temperature ($T_{\rm ex}$) estimated from the HDCO lines. This is based on the assumption that the three species probe the same component, an hypothesis reinforced by the similarity of the extracted line profiles at each position.
As reported in Tab. \ref{temperatures}, the HDCO column density towards VLA1623 A is 5$\pm$3 $\times$ 10$^{13}$ cm$^{-2}$, an order of magnitude higher than that measured in the VLA1623 B red-shifted and SE cavities, $N_{\rm HDCO}$ $\simeq$
5--6 $\times$ 10$^{12}$ cm$^{-2}$. The HDCO column densities in the streamers II and VII are $\geq$ 10$^{12}$ cm$^{-2}$.
The formaldehyde column densities are $\simeq$ 3 $\times$ 10$^{14}$ cm$^{-2}$ (VLA1623 A), and 3--8 $\times$ 10$^{13}$ cm$^{-2}$ for the
other positions.
Finally, the D$_2$CO column densities are $\geq$ 10$^{12}$ cm$^{-2}$ in
all the positions listed in Tab. \ref{flows}.

Table \ref{temperatures} reports, for each position, the abundance ratios between the deuterated isotopologue and the main species, H$_2$CO, which are equal to the column densities ratios.
Then, following \citet{Manigand2019}, the deuterium fractions, D/H, are obtained from the abundance ratios: D/H = 0.5 $\times$ $N_{\rm HDCO}$/$N_{\rm H_2CO}$, and D/H = $\sqrt{N_{\rm D_2CO}/N_{\rm H_2CO}}$. 

The D-fraction of H$_{2}$CO in the inner envelope of VLA1623 A is 8\% for HDCO, and
6\% for D$_2$CO.
Remarkably, we report the first estimate of D/H in the cavities opened by the protostellar outflow. Namely, D/H is  6\% (HDCO), and 
35\% (D$_2$CO) in the SE outflow cavity; and 4\% (HDCO), and 
$\leq$ 22\% (D$_2$CO) in the NW outflow cavity.
Finally, the estimate of D/H in the streamers is not accurate because we only have an upper limit of the excitation temperature, hence a lower limit of the column densities of the three species. 
However, assuming that H$_2$CO, HDCO, and D$_2$CO are excited at the inferred $T_{\rm ex}$ upper limit, we obtain an estimate of the D/H in the streamers: D/H is 3\%--15\% (for HDCO), and 17\% 
(for D$_2$CO).

We show the intensity ratio maps of HDCO/H$_2$CO and D$_2$CO/H$_2$CO in Fig. \ref{RatioMaps}. The maps are integrated taking into account the velocity ranges imaging the V (orthern blue-shifted streamer towards A), VII (outflow cavities driven by A ), and VII (southern red-shifted streamer towards A) components listed in Table \ref{flows} to be consistent with the deuteration derived in the system. The distribution of the ratio maps is consistent with that found in Table \ref{temperatures}.

\subsection{Comparison with previous results on H$_2$CO deuteration}
\label{comparison}

\begin{figure*}
\includegraphics[width=13cm]{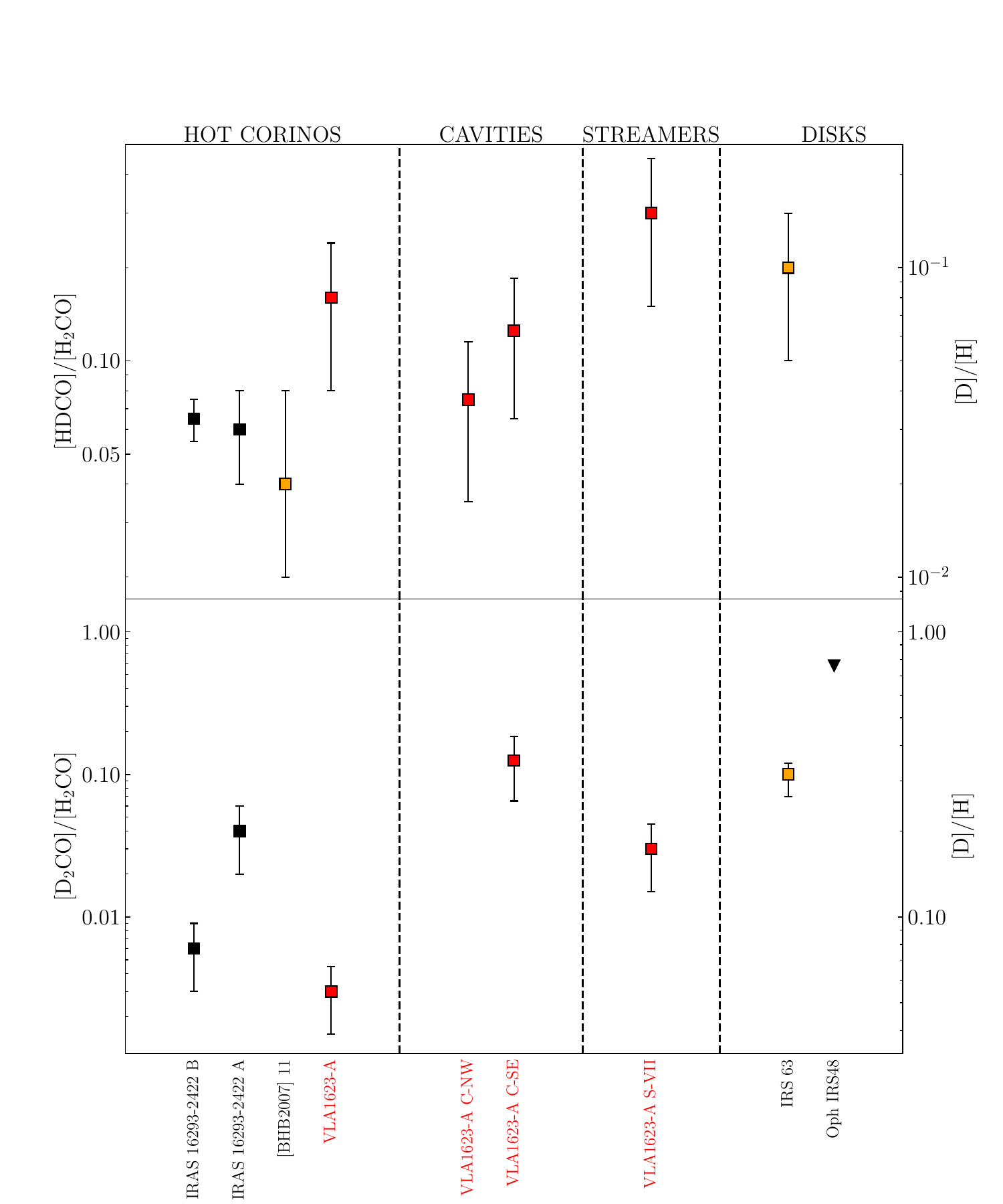}
\vspace{+2cm}
\caption{Abundance ratios [HDCO]/[H$_2$CO] ({\it upper panel}), and [D$_2$CO]/[H$_2$CO] ({\it lower panel}) estimated through interferometric observations. The abundance ratios (left y-axis) and D/H (right y-axis) derived for VLA1623 hot corino (VLA1623-A), NW and SE outflow cavities (VLA1623-A C-NW and C-SE), and the streamer VII (VLA1623-A S-VII) are indicated by red squares (See Table \ref{temperatures}). Previous estimates in Class 0 hot corinos and disks from the literature are reported with black squares \citep{Persson2018,manigand20,vandermarel21b, Brunken2022}, and with orange ones for the ALMA-FAUST studies \citep{Evans2023, Podio2024}. Upper limits are marked with triangles. The source names are labeled below the bottom panel (in red for the values estimated in this paper for VLA1623). The axes are logarithmic.}
\label{deut_vla1623}
\end{figure*}

Figure \ref{deut_vla1623} compares the D/H values estimated for VLA1623 with previous interferometric measurements obtained for Class 0 hot-corinos and disks either in the context of the FAUST LP \citep{Evans2023, Podio2024} or in previous ALMA studies \citep{Persson2018, Manigand2020,vandermarel21b, Brunken2022}.
The range of D/H values found in the envelope, outflow cavities, and infalling streamer of VLA1623 A agrees with earlier findings.
Interestingly, both the [HDCO]/[H$_2$CO] and [D$_2$CO]/[H$_2$CO] abundance ratios in the streamer are in agreement with those found by \citet{Podio2024} in the Class I disk IRS~63. They report a significant increase of the abundance of D$_2$CO  at the location where the infalling streamer collides with the disk, causing a shock and leading to the release in gas-phase of the molecules trapped in the dust mantles. The accretion shock thus reveal the composition of the ices in the disk of IRS~63. 
However, they cannot determine if the ice composition is inherited from the prestellar stage, or if ices are formed and/or reprocessed in the disk itself.
Our observations of VLA1623 shows D$_2$CO-enriched mantles in the streamer, with D/H ratios similar to those in the Class I disk of IRS~63, thus suggesting that deuteration is mostly set at the prestellar stage, and inherited by the disk at its formation and/or through streamers. The comparison within each class of regions and between types of regions is difficult because of the inherent differences that are already known. These variations can be driven by local physical conditions. Hence the present data call for further observations of other star-forming regions to look for differences in D/H within the same structure (such as along streamers or cavities) which may share similar conditions such as timescale, evolution, and physical conditions.


\begin{table*}
\centering
     \begin{tabular}{lcccccc}
     \hline
Species&$T_{\rm ex}$& Velocity range &Intensity &$N$ &[X]/[H$_{2}$CO] & D/H \\
 & (K)  & (km s$^{-1}$) & (K km s$^{-1}$) &(cm$^{-2}$)& & $\% $  \\
\hline 
\multicolumn{7}{c}{VLA1623 A}\\ 
\hline 
H$_2$CO &$125$ &--1.0 -- +2.0  &5.9$\pm$0.2 & $3\times10^{14}$ & --&--\\
HDCO  &$125 (60)$ &--1.0 -- +2.0& 6.1$\pm$0.2 (38K) / 4.0$\pm$0.1 (63K)
&$5(3)\times10^{13}$ &0.160 & 8 \\
D$_2$CO &$125$ &--1.0 -- +2.0 & 0.4$\pm$0.1
&$1\times10^{12}$ & 0.003& 6 \\
\hline
\multicolumn{7}{c}{NW Outflow Cavity (VLA1623 B position)}\\ 
\hline 
H$_2$CO &$28$  & +5.0 -- +13.0& 8.6$\pm$0.2&$8\times10^{13}$ &-- &-- \\
HDCO  &$28 (12)$  & +5.0 -- +13.0& 0.9 (38K) / 0.3 (63K)&$6(4)\times10^{12}$ & 0.075& 4 \\
D$_2$CO & $28$& +5.0 -- +13.0& $\le0.7\pm$0.1& $\leq$ $4\times10^{12}$ & $\leq$ 0.050 & $\leq$ 22 \\
\hline
\multicolumn{7}{c}{SE Outflow Cavity, VI}\\
\hline 
H$_2$CO &$37$  & +3.1 -- +5.0& 3.7$\pm$0.1&$4\times10^{13}$ &-- & --\\
HDCO  & $37 (19)$  &+3.1 -- +5.0 & 0.7 (38K) / 0.3 (63K)&$5(4)\times10^{12}$ & 0.125 & 6 \\
D$_2$CO & $37$&+3.1 -- +5.0 & 1.1$\pm$0.1& $5\times10^{12}$ & 0.125 & 35 \\
\hline
\multicolumn{7}{c}{Streamer II (SE Outflow Cavity position)}\\
\hline 
H$_2$CO & $\leq$ 15 & +0.8 -- +2.1&3.8$\pm$0.1 & $\geq$ $3\times10^{13}$  & -- & --\\
HDCO  & $\leq$ 15 & +0.8 -- +2.1 &0.7$\pm$0.1 (38 K) & $\geq$ $2\times10^{12}$ & 0.067 & 3 \\
D$_2$CO & -- & --  & -- & --& -- & -- \\
\hline
\multicolumn{7}{c}{Streamer VII}\\
\hline 
H$_2$CO & $\leq$ 10 & +2.5 -- +6.1 &4.1$\pm$0.1 & $\geq$ $3\times10^{13}$&-- & --\\
HDCO  & $\leq$ $10$ & +2.5 -- +6.1 & 0.5$\pm$0.1 (38 K)& $\geq$ $9\times10^{12}$ &0.300 &15 \\
D$_2$CO &  $\leq$ 10  &+2.5 -- +6.1 & 1.4$\pm$0.1&$\geq$ $1\times10^{12}$ & 0.030& 17\\
\hline
     \end{tabular}
     
\caption{Velocity ranges used for line integration (see Fig. \ref{superspectra}), integrated line intensities, excitation temperature ($T_{\rm ex}$), molecular column densities ($N$), abundance ratios, and D/H ratios towards the protostellar continuum peaks, and selected positions representing the outflow cavities, and the streamers. For HDCO we report the intensity of both lines, indicating their $E_{\rm up}$ (38 K, and 63 K) in parenthesis next to the intensity. The excitation temperature is derived from the ratio of the HDCO lines. For H$_2$CO and D$_2$CO, $T_{\rm ex}$ is assumed to be equal to that derived from HDCO. For the streamers, we report the abundance ratios and D/H derived from the lower limits of the column densities.
} 
\label{temperatures}
\end{table*}

\subsection{On the origin of D$_2$CO extended emission}
\label{D1}

ALMA-FAUST interferometric observations show extended emission of D$_{2}$CO toward the multiple system VLA1623-2417 for the first time. 
As reported in Sect. \ref{results}, D$_{2}$CO emission reveals a puzzling spatial distribution, which does not match that of H$_2$CO and HDCO.
While H$_2$CO and HDCO emissions have their peak in the hot corino toward VLA1623 A, in the outflow cavities, and in the redshifted southern streamer, D$_2$CO  probes mostly the flattened envelope perpendicular to the outflow direction (see Fig. \ref{Mom0MAPS}), which is fainter in H$_2$CO and HDCO.
A different spatial distribution of D$_2$CO with respect to that of H$_2$CO and HDCO has also been found towards another FAUST protostellar source, the disk-streamer system of IRS~63, and has been interpreted as due to a different chemical origin of D$_2$CO, with respect to H$_2$CO and HDCO \citep{Podio2024}.

More specifically, formaldehyde and its deuterated isotopologues can form both in the gas phase and on the surface of dust grains.
In warm gas \citep[$\sim$ 50 K, ][]{Parise2009} H$_2$CO, HDCO, and D$_2$CO follow a similar formation path, starting from CH$_3^+$ which reacts either with H$_2$ (to form H$_2$CO), or with HD (to form CH$_2$D$^+$ and CHD$_2^+$, and from the latter HDCO and D$_2$CO) \citep[e.g., ][]{Roueff2007, Roueff2013-CH2D+}.
However, the different spatial distribution of D$_2$CO points to its formation on the surface of dust grains, following the scheme presented in Fig. 4 by \citet{Podio2024}.
The processes occurring on the grains may cause an enhancement of the D$_2$CO abundance with respect to that of HDCO and H$_2$CO on the ices.
Then, when the ice mantles are released in the gas phase due to shocks (as in the case of IRS~63) or to a radiation field, we may observe an enhancement of D$_2$CO emission, with respect to HDCO and H$_2$CO.

Following the above reasoning, we propose that in the case of the VLA1623 protostellar cluster, the fact that D$_2$CO is detected in an extended region where HDCO and H$_2$CO emission is faint, is due to deuterated molecules formation on the grains in the cold prestellar phase and release in the gas phase due to irradiation within a dense photodissociation region (PDR).





Figure \ref{D2CO_Mom1and2} shows the moment 2 (velocity dispersion) map of D$_2$CO (4$_{0,4}$ -- 3$_{0,3}$) emission, which indicates that the extended emission partially traces the outflow cavities and the  streamer as listed in Table \ref{flows}. The extended D$_2$CO emission shows a narrow velocity dispersion between 0.3 and 1.2 km s$^{-1}$. 
This is not consistent with the release of D$_2$CO molecules in shocks, which would produce large line profiles. Thus, we investigate whether the observed D$_2$CO emission can be caused by the release of the icy mantles in gas-phase due to a PDR. 
\citet{Rawlings2013} presented infrared observations  taken with WISE (Wide-field Infrared Survey Explorer), {\it Herschel}, and 2MASS (Two
Micron All-Sky Survey), and visual extinction maps towards the VLA1623-2417 region in the Ophiuchus cloud.
These maps show that VLA1623-2417 is illuminated by two nearby B stars (HD147889 and $\rho$ Oph AB). Their study suggests high irradiation and an increased emission of PAHs (polycyclic aromatic hydrocarbons). The 2MASS catalog also indicate high visual extinction ($A_{\rm v}$ > 30 mag) towards VLA1623-2417.
In addition, polarisation studies with JCMT SCUBA by \citet{Pattle2019} showed evidence of an external radiation field which may cause dust grain alignment. 

In summary, the VLA1623-2417 protostellar cluster could be affected by a locally stronger radiation field. Thus, the narrow velocity dispersion of D$_2$CO emission (Fig. \ref{D2CO_Mom1and2}) could trace the front of the radiation field caused by the massive B stars and OB association in the Oph A core. If this is the case, D$_2$CO molecules are photo-desorbed from the grains. 
\section{Summary and Conclusions}
\label{conclusions}

The FAUST ALMA Large Program has imaged deuterated formaldehyde emission in the VLA1623-2417 protostellar cluster down to the spatial scale of 50\,au. Emission from H$_2$CO (3$_{0,3}$ -- 2$_{0,2}$), HDCO (4$_{1,4}$ -- 3$_{1,3}$), HDCO (4$_{2,2}$ -- 3$_{2,1}$), D$_2$CO (4$_{0,4}$ -- 3$_{0,3}$), and D$_2$CO (4$_{2,3}$ -- 3$_{2,2}$) has been detected and shows both extended and compact structures. The main findings are summarised as follows:

\begin{enumerate}

\item
The binary system VLA1623 A is associated with formaldehyde and its deuterated isotopologues. Conversely, the VLA1623 B and W protostars have been detected in H$_2$CO but not in HDCO and D$_2$CO.
High-velocity formaldehyde emissions trace the two counter-rotating, spatially unresolved cores ($\leq$ 50 au) of VLA1623 A and  VLA1623 B, and the brightest blue-shifted side of the envelope of VLA1623 W;

\item 
H$_2$CO, HDCO, and D$_2$CO also probe extended structures, including the symmetric and rotating cavities created by the VLA1623 A outflow, and two extended (approximately 1000 au) accretion streamers feeding the VLA1623 A+B system. 

\item
Based on the ratio of the HDCO lines, we estimate the gas temperature to be approximately 125 K towards the VLA1623 A hot-corino, between 20-40 K in the cavities, and $\leq$ 15 K in the streamers;

\item 
D$_2$CO reveals an extended component, which is fainter in H$_2$CO and HDCO, that probes the flattened envelope perpendicular to the outflow direction. We suggest that the extended D$_2$CO emission results from D$_2$CO formation on grains during the cold prestellar phase and its release into the gas phase due to the photo-dissociation region caused by two nearby B stars (HD147889 and $\rho$ Oph AB);

\item
The D-fraction of formaldehyde in the hot-corino, outflow cavities, and infalling streamers of VLA1623 A lies in the following ranges: [HDCO]/[H$_2$CO] $\sim 0.07-0.3$, and [D$_2$CO]/[H$_2$CO] $\sim 0.003-0.13$. These values are comparable to those previously obtained for other Class 0 hot-corinos, as well as for Class I and II disks. 
    
\end{enumerate}

Our results support the hypothesis that deuterated species primarily form during the cold prestellar stage and are subsequently inherited by the forming protostar.

\section*{Acknowledgements}
We are grateful to the referee P.T.P. Ho for his instructive report. We also thank N. Balucani for precious discussions on the chemistry of formaldehyde isotopologues. 
This project has received funding from the EC H2020 research and innovation
programme for: (i) the project "Astro-Chemical Origins” (ACO, No 811312), and (ii) the European Research Council (ERC) project “The Dawn of Organic Chemistry” (DOC, No 741002).  
LP, GS, and ClCo acknowledge the PRIN-MUR 2020 BEYOND-2p (Astrochemistry beyond the second period elements, Prot. 2020AFB3FX), the project ASI-Astrobiologia 2023 MIGLIORA (Modeling Chemical Complexity, F83C23000800005), the INAF-GO 2023 fundings PROTO-SKA (Exploiting ALMA data to study planet forming disks: preparing the advent of SKA, C13C23000770005), the INAF Mini-Grant 2022 “Chemical Origins” (PI: L. Podio), and financial support under the National Recovery and Resilience Plan (NRRP), Mission 4, Component 2, Investment 1.1, Call for tender No. 104 published on 2.2.2022 by the Italian Ministry of University and Research (MUR), funded by the European Union – NextGenerationEU– Project Title 2022JC2Y93 Chemical Origins: linking the fossil composition of the Solar System with the chemistry of protoplanetary disks – CUP J53D23001600006 - Grant Assignment Decree No. 962 adopted on 30.06.2023 by the Italian Ministry of University and Research (MUR). G.S. also acknowledges the INAF-Minigrant 2023 TRIESTE ("TRacing the chemIcal hEritage of our originS: from proTostars to planEts"; PI: G. Sabatini). D.J. is supported by NRC Canada and by an NSERC Discovery Grant.
This paper makes use of the following ALMA data: ADS/JAO.ALMA\#2018.1.01205.L. ALMA is a partnership of ESO (representing its member states), NSF (USA), and NINS (Japan), together with NRC (Canada), MOST and ASIAA (Taiwan), and KASI (Republic of Korea), in cooperation with the Republic of Chile. The Joint ALMA Observatory is operated by ESO, AUI/NRAO and NAOJ. The National Radio Astronomy Observatory is a facility of the National Science Foundation operated under a cooperative agreement by Associated Universities, Inc. \\

\section*{Data Availability}
The raw data are available on the ALMA archive at the end of the proprietary period (ADS/JAO.ALMA\#2018.1.01205.L).



\bibliographystyle{mnras}
\bibliography{example} 

\begin{thebibliography}{}
\makeatletter
\relax
\def\mn@urlcharsother{\let\do\@makeother \do\$\do\&\do\#\do\^\do\_\do\%\do\~}
\def\mn@doi{\begingroup\mn@urlcharsother \@ifnextchar [ {\mn@doi@} {\mn@doi@[]}}
\def\mn@doi@[#1]#2{\def\@tempa{#1}\ifx\@tempa\@empty \href {http://dx.doi.org/#2} {doi:#2}\else \href {http://dx.doi.org/#2} {#1}\fi \endgroup}
\def\mn@eprint#1#2{\mn@eprint@#1:#2::\@nil}
\def\mn@eprint@arXiv#1{\href {http://arxiv.org/abs/#1} {{\tt arXiv:#1}}}
\def\mn@eprint@dblp#1{\href {http://dblp.uni-trier.de/rec/bibtex/#1.xml} {dblp:#1}}
\def\mn@eprint@#1:#2:#3:#4\@nil{\def\@tempa {#1}\def\@tempb {#2}\def\@tempc {#3}\ifx \@tempc \@empty \let \@tempc \@tempb \let \@tempb \@tempa \fi \ifx \@tempb \@empty \def\@tempb {arXiv}\fi \@ifundefined {mn@eprint@\@tempb}{\@tempb:\@tempc}{\expandafter \expandafter \csname mn@eprint@\@tempb\endcsname \expandafter{\@tempc}}}

\bibitem[\protect\citeauthoryear{{Andre}, {Martin-Pintado}, {Despois}  \& {Montmerle}}{{Andre} et~al.}{1990}]{Andre1990}
{Andre} P.,  {Martin-Pintado} J.,  {Despois} D.,   {Montmerle} T.,  1990, \aap, \href {https://ui.adsabs.harvard.edu/abs/1990A&A...236..180A} {236, 180}

\bibitem[\protect\citeauthoryear{{Bacmann}, {Lefloch}, {Ceccarelli}, {Castets}, {Steinacker}  \& {Loinard}}{{Bacmann} et~al.}{2002}]{Bacmann2002}
{Bacmann} A.,  {Lefloch} B.,  {Ceccarelli} C.,  {Castets} A.,  {Steinacker} J.,   {Loinard} L.,  2002, \mn@doi [\aap] {10.1051/0004-6361:20020652}, \href {https://ui.adsabs.harvard.edu/abs/2002A&A...389L...6B} {389, L6}

\bibitem[\protect\citeauthoryear{{Bacmann}, {Lefloch}, {Ceccarelli}, {Steinacker}, {Castets}  \& {Loinard}}{{Bacmann} et~al.}{2003}]{Bacmann2003}
{Bacmann} A.,  {Lefloch} B.,  {Ceccarelli} C.,  {Steinacker} J.,  {Castets} A.,   {Loinard} L.,  2003, \mn@doi [\apjl] {10.1086/374263}, \href {https://ui.adsabs.harvard.edu/abs/2003ApJ...585L..55B} {585, L55}

\bibitem[\protect\citeauthoryear{{Bianchi} et~al.,}{{Bianchi} et~al.}{2017}]{Bianchi2017}
{Bianchi} E.,  et~al., 2017, \mn@doi [\mnras] {10.1093/mnras/stx252}, \href {https://ui.adsabs.harvard.edu/abs/2017MNRAS.467.3011B} {467, 3011}

\bibitem[\protect\citeauthoryear{{Brunken}, {Booth}, {Leemker}, {Nazari}, {van der Marel}  \& {van Dishoeck}}{{Brunken} et~al.}{2022}]{Brunken2022}
{Brunken} N. G.~C.,  {Booth} A.~S.,  {Leemker} M.,  {Nazari} P.,  {van der Marel} N.,   {van Dishoeck} E.~F.,  2022, \mn@doi [\aap] {10.1051/0004-6361/202142981}, \href {https://ui.adsabs.harvard.edu/abs/2022A&A...659A..29B} {659, A29}

\bibitem[\protect\citeauthoryear{{CASA Team} et~al.,}{{CASA Team} et~al.}{2022}]{CASA2022}
{CASA Team} et~al., 2022, \mn@doi [\pasp] {10.1088/1538-3873/ac9642}, \href {https://ui.adsabs.harvard.edu/abs/2022PASP..134k4501C} {134, 114501}

\bibitem[\protect\citeauthoryear{{Caselli} \& {Ceccarelli}}{{Caselli} \& {Ceccarelli}}{2012}]{Caselli2012}
{Caselli} P.,  {Ceccarelli} C.,  2012, \mn@doi [\aapr] {10.1007/s00159-012-0056-x}, \href {https://ui.adsabs.harvard.edu/abs/2012A&ARv..20...56C} {20, 56}

\bibitem[\protect\citeauthoryear{{Caselli}, {Walmsley}, {Tafalla}, {Dore}  \& {Myers}}{{Caselli} et~al.}{1999}]{Caselli1999}
{Caselli} P.,  {Walmsley} C.~M.,  {Tafalla} M.,  {Dore} L.,   {Myers} P.~C.,  1999, \mn@doi [\apjl] {10.1086/312280}, \href {https://ui.adsabs.harvard.edu/abs/1999ApJ...523L.165C} {523, L165}

\bibitem[\protect\citeauthoryear{{Caselli}, {Walmsley}, {Zucconi}, {Tafalla}, {Dore}  \& {Myers}}{{Caselli} et~al.}{2002}]{Caselli2002}
{Caselli} P.,  {Walmsley} C.~M.,  {Zucconi} A.,  {Tafalla} M.,  {Dore} L.,   {Myers} P.~C.,  2002, \mn@doi [\apj] {10.1086/324302}, \href {https://ui.adsabs.harvard.edu/abs/2002ApJ...565..344C} {565, 344}

\bibitem[\protect\citeauthoryear{{Ceccarelli}, {Caselli}, {Herbst}, {Tielens}  \& {Caux}}{{Ceccarelli} et~al.}{2007}]{Ceccarelli2007}
{Ceccarelli} C.,  {Caselli} P.,  {Herbst} E.,  {Tielens} A.~G.~G.~M.,   {Caux} E.,  2007, in {Reipurth} B.,  {Jewitt} D.,   {Keil} K.,  eds, Protostars and Planets V. p.~47 (\mn@eprint {arXiv} {astro-ph/0603018})

\bibitem[\protect\citeauthoryear{{Ceccarelli}, {Caselli}, {Bockel{\'e}e-Morvan}, {Mousis}, {Pizzarello}, {Robert}  \& {Semenov}}{{Ceccarelli} et~al.}{2014}]{Ceccarelli2014}
{Ceccarelli} C.,  {Caselli} P.,  {Bockel{\'e}e-Morvan} D.,  {Mousis} O.,  {Pizzarello} S.,  {Robert} F.,   {Semenov} D.,  2014, in {Beuther} H.,  {Klessen} R.~S.,  {Dullemond} C.~P.,   {Henning} T.,  eds, Protostars and Planets VI. pp 859--882 (\mn@eprint {arXiv} {1403.7143}), \mn@doi{10.2458/azu_uapress_9780816531240-ch037}

\bibitem[\protect\citeauthoryear{{Ceccarelli} et~al.,}{{Ceccarelli} et~al.}{2023}]{Ceccarelli2023}
{Ceccarelli} C.,  et~al., 2023, in {Inutsuka} S.,  {Aikawa} Y.,  {Muto} T.,  {Tomida} K.,   {Tamura} M.,  eds,  Astronomical Society of the Pacific Conference Series Vol. 534, Astronomical Society of the Pacific Conference Series. p.~379

\bibitem[\protect\citeauthoryear{{Chac{\'o}n-Tanarro} et~al.,}{{Chac{\'o}n-Tanarro} et~al.}{2019}]{Chacon2019}
{Chac{\'o}n-Tanarro} A.,  et~al., 2019, \mn@doi [\aap] {10.1051/0004-6361/201832703}, \href {https://ui.adsabs.harvard.edu/abs/2019A&A...622A.141C} {622, A141}

\bibitem[\protect\citeauthoryear{{Chahine} et~al.,}{{Chahine} et~al.}{2024}]{Chahine2024}
{Chahine} L.,  et~al., 2024, \mn@doi [\mnras] {10.1093/mnrasl/slae080}, \href {https://ui.adsabs.harvard.edu/abs/2024MNRAS.534L..48C} {534, L48}

\bibitem[\protect\citeauthoryear{{Codella}, {Ceccarelli}, {Chandler}, {Sakai}, {Yamamoto}  \& {FAUST Team}}{{Codella} et~al.}{2021}]{Codella2021}
{Codella} C.,  {Ceccarelli} C.,  {Chandler} C.,  {Sakai} N.,  {Yamamoto} S.,   {FAUST Team} 2021, \mn@doi [Frontiers in Astronomy and Space Sciences] {10.3389/fspas.2021.782006}, \href {https://ui.adsabs.harvard.edu/abs/2021FrASS...8..227C} {8, 227}

\bibitem[\protect\citeauthoryear{{Codella} et~al.,}{{Codella} et~al.}{2022}]{Codella2022}
{Codella} C.,  et~al., 2022, \mn@doi [\mnras] {10.1093/mnras/stac1802}, \href {https://ui.adsabs.harvard.edu/abs/2022MNRAS.515..543C} {515, 543}

\bibitem[\protect\citeauthoryear{{Codella} et~al.,}{{Codella} et~al.}{2024}]{Codella2024}
{Codella} C.,  et~al., 2024, arXiv e-prints, \href {https://ui.adsabs.harvard.edu/abs/2024arXiv240210258C} {p. arXiv:2402.10258}

\bibitem[\protect\citeauthoryear{{Cooke}, {Pettini}  \& {Steidel}}{{Cooke} et~al.}{2018}]{Cooke2018}
{Cooke} R.~J.,  {Pettini} M.,   {Steidel} C.~C.,  2018, \mn@doi [\apj] {10.3847/1538-4357/aaab53}, \href {https://ui.adsabs.harvard.edu/abs/2018ApJ...855..102C} {855, 102}

\bibitem[\protect\citeauthoryear{{Crapsi}, {Caselli}, {Walmsley}, {Myers}, {Tafalla}, {Lee}  \& {Bourke}}{{Crapsi} et~al.}{2005}]{Crapsi2005}
{Crapsi} A.,  {Caselli} P.,  {Walmsley} C.~M.,  {Myers} P.~C.,  {Tafalla} M.,  {Lee} C.~W.,   {Bourke} T.~L.,  2005, \mn@doi [\apj] {10.1086/426472}, \href {https://ui.adsabs.harvard.edu/abs/2005ApJ...619..379C} {619, 379}

\bibitem[\protect\citeauthoryear{{Drozdovskaya} et~al.,}{{Drozdovskaya} et~al.}{2021}]{Drozdovskaya2021}
{Drozdovskaya} M.~N.,  et~al., 2021, \mn@doi [\mnras] {10.1093/mnras/staa3387}, \href {https://ui.adsabs.harvard.edu/abs/2021MNRAS.500.4901D} {500, 4901}

\bibitem[\protect\citeauthoryear{{Emprechtinger}, {Caselli}, {Volgenau}, {Stutzki}  \& {Wiedner}}{{Emprechtinger} et~al.}{2009}]{Emprechtinger2009}
{Emprechtinger} M.,  {Caselli} P.,  {Volgenau} N.~H.,  {Stutzki} J.,   {Wiedner} M.~C.,  2009, \mn@doi [\aap] {10.1051/0004-6361:200810324}, \href {https://ui.adsabs.harvard.edu/abs/2009A&A...493...89E} {493, 89}

\bibitem[\protect\citeauthoryear{{Evans} et~al.,}{{Evans} et~al.}{2023}]{Evans2023}
{Evans} L.,  et~al., 2023, \mn@doi [\aap] {10.1051/0004-6361/202346428}, \href {https://ui.adsabs.harvard.edu/abs/2023A&A...678A.160E} {678, A160}

\bibitem[\protect\citeauthoryear{{Fuchs}, {Cuppen}, {Ioppolo}, {Romanzin}, {Bisschop}, {Andersson}, {van Dishoeck}  \& {Linnartz}}{{Fuchs} et~al.}{2009}]{Fuchs2009}
{Fuchs} G.~W.,  {Cuppen} H.~M.,  {Ioppolo} S.,  {Romanzin} C.,  {Bisschop} S.~E.,  {Andersson} S.,  {van Dishoeck} E.~F.,   {Linnartz} H.,  2009, \mn@doi [\aap] {10.1051/0004-6361/200810784}, \href {https://ui.adsabs.harvard.edu/abs/2009A&A...505..629F} {505, 629}

\bibitem[\protect\citeauthoryear{{Gagn{\'e}} et~al.,}{{Gagn{\'e}} et~al.}{2018}]{Gagne2018}
{Gagn{\'e}} J.,  et~al., 2018, \mn@doi [\apj] {10.3847/1538-4357/aaae09}, \href {https://ui.adsabs.harvard.edu/abs/2018ApJ...856...23G} {856, 23}

\bibitem[\protect\citeauthoryear{{Garrod}, {Jin}, {Matis}, {Jones}, {Willis}  \& {Herbst}}{{Garrod} et~al.}{2022}]{Garrod2022}
{Garrod} R.~T.,  {Jin} M.,  {Matis} K.~A.,  {Jones} D.,  {Willis} E.~R.,   {Herbst} E.,  2022, \mn@doi [\apjs] {10.3847/1538-4365/ac3131}, \href {https://ui.adsabs.harvard.edu/abs/2022ApJS..259....1G} {259, 1}

\bibitem[\protect\citeauthoryear{{Gerlich}, {Herbst}  \& {Roueff}}{{Gerlich} et~al.}{2002}]{Gerlich2002}
{Gerlich} D.,  {Herbst} E.,   {Roueff} E.,  2002, \mn@doi [\planss] {10.1016/S0032-0633(02)00094-6}, \href {https://ui.adsabs.harvard.edu/abs/2002P&SS...50.1275G} {50, 1275}

\bibitem[\protect\citeauthoryear{{Hara} et~al.,}{{Hara} et~al.}{2021}]{Hara2021}
{Hara} C.,  et~al., 2021, \mn@doi [\apj] {10.3847/1538-4357/abb810}, \href {https://ui.adsabs.harvard.edu/abs/2021ApJ...912...34H} {912, 34}

\bibitem[\protect\citeauthoryear{{Harris} et~al.,}{{Harris} et~al.}{2018}]{Harris2018}
{Harris} R.~J.,  et~al., 2018, \mn@doi [\apj] {10.3847/1538-4357/aac6ec}, \href {https://ui.adsabs.harvard.edu/abs/2018ApJ...861...91H} {861, 91}

\bibitem[\protect\citeauthoryear{{Hsieh}, {Lai}, {Cheong}, {Ko}, {Li}  \& {Murillo}}{{Hsieh} et~al.}{2020}]{Hsieh2020}
{Hsieh} C.-H.,  {Lai} S.-P.,  {Cheong} P.-I.,  {Ko} C.-L.,  {Li} Z.-Y.,   {Murillo} N.~M.,  2020, \mn@doi [\apj] {10.3847/1538-4357/ab7b69}, \href {https://ui.adsabs.harvard.edu/abs/2020ApJ...894...23H} {894, 23}

\bibitem[\protect\citeauthoryear{{Loinard}, {Castets}, {Ceccarelli}, {Caux}  \& {Tielens}}{{Loinard} et~al.}{2001}]{Loinard2001}
{Loinard} L.,  {Castets} A.,  {Ceccarelli} C.,  {Caux} E.,   {Tielens} A.~G.~G.~M.,  2001, \mn@doi [\apjl] {10.1086/320331}, \href {https://ui.adsabs.harvard.edu/abs/2001ApJ...552L.163L} {552, L163}

\bibitem[\protect\citeauthoryear{{Looney}, {Mundy}  \& {Welch}}{{Looney} et~al.}{2000}]{Looney2000}
{Looney} L.~W.,  {Mundy} L.~G.,   {Welch} W.~J.,  2000, \mn@doi [\apj] {10.1086/308239}, \href {http://adsabs.harvard.edu/abs/2000ApJ...529..477L} {529, 477}

\bibitem[\protect\citeauthoryear{{Manigand} et~al.,}{{Manigand} et~al.}{2019}]{Manigand2019}
{Manigand} S.,  et~al., 2019, \mn@doi [\aap] {10.1051/0004-6361/201832844}, \href {https://ui.adsabs.harvard.edu/abs/2019A&A...623A..69M} {623, A69}

\bibitem[\protect\citeauthoryear{{Manigand} et~al.,}{{Manigand} et~al.}{2020a}]{manigand20}
{Manigand} S.,  et~al., 2020a, \mn@doi [\aap] {10.1051/0004-6361/201936299}, \href {https://ui.adsabs.harvard.edu/abs/2020A&A...635A..48M} {635, A48}

\bibitem[\protect\citeauthoryear{{Manigand} et~al.,}{{Manigand} et~al.}{2020b}]{Manigand2020}
{Manigand} S.,  et~al., 2020b, \mn@doi [\aap] {10.1051/0004-6361/201936299}, \href {https://ui.adsabs.harvard.edu/abs/2020A&A...635A..48M} {635, A48}

\bibitem[\protect\citeauthoryear{{Mercimek} et~al.,}{{Mercimek} et~al.}{2022}]{Mercimek2022}
{Mercimek} S.,  et~al., 2022, \mn@doi [\aap] {10.1051/0004-6361/202141790}, \href {https://ui.adsabs.harvard.edu/abs/2022A&A...659A..67M} {659, A67}

\bibitem[\protect\citeauthoryear{{Mercimek} et~al.,}{{Mercimek} et~al.}{2023}]{Mercimek2023}
{Mercimek} S.,  et~al., 2023, \mn@doi [\mnras] {10.1093/mnras/stad964}, \href {https://ui.adsabs.harvard.edu/abs/2023MNRAS.522.2384M} {522, 2384}

\bibitem[\protect\citeauthoryear{{M{\"u}ller}, {Schl{\"o}der}, {Stutzki}  \& {Winnewisser}}{{M{\"u}ller} et~al.}{2005}]{Muller2005}
{M{\"u}ller} H. S.~P.,  {Schl{\"o}der} F.,  {Stutzki} J.,   {Winnewisser} G.,  2005, \mn@doi [Journal of Molecular Structure] {10.1016/j.molstruc.2005.01.027}, \href {https://ui.adsabs.harvard.edu/abs/2005JMoSt.742..215M} {742, 215}

\bibitem[\protect\citeauthoryear{{Murillo}, {Lai}, {Bruderer}, {Harsono}  \& {van Dishoeck}}{{Murillo} et~al.}{2013}]{Murillo2013disk}
{Murillo} N.~M.,  {Lai} S.-P.,  {Bruderer} S.,  {Harsono} D.,   {van Dishoeck} E.~F.,  2013, \mn@doi [\aap] {10.1051/0004-6361/201322537}, \href {https://ui.adsabs.harvard.edu/abs/2013A&A...560A.103M} {560, A103}

\bibitem[\protect\citeauthoryear{{Murillo}, {Harsono}, {McClure}, {Lai}  \& {Hogerheijde}}{{Murillo} et~al.}{2018a}]{Murillo2018L}
{Murillo} N.~M.,  {Harsono} D.,  {McClure} M.,  {Lai} S.~P.,   {Hogerheijde} M.~R.,  2018a, \mn@doi [\aap] {10.1051/0004-6361/201833420}, \href {https://ui.adsabs.harvard.edu/abs/2018A&A...615L..14M} {615, L14}

\bibitem[\protect\citeauthoryear{{Murillo}, {van Dishoeck}, {van der Wiel}, {J{\o}rgensen}, {Drozdovskaya}, {Calcutt}  \& {Harsono}}{{Murillo} et~al.}{2018b}]{Murillo2018}
{Murillo} N.~M.,  {van Dishoeck} E.~F.,  {van der Wiel} M.~H.~D.,  {J{\o}rgensen} J.~K.,  {Drozdovskaya} M.~N.,  {Calcutt} H.,   {Harsono} D.,  2018b, \mn@doi [\aap] {10.1051/0004-6361/201731724}, \href {https://ui.adsabs.harvard.edu/abs/2018A&A...617A.120M} {617, A120}

\bibitem[\protect\citeauthoryear{{Ohashi} et~al.,}{{Ohashi} et~al.}{2022}]{Ohashi2022}
{Ohashi} S.,  et~al., 2022, \mn@doi [\apj] {10.3847/1538-4357/ac4cae}, \href {https://ui.adsabs.harvard.edu/abs/2022ApJ...927...54O} {927, 54}

\bibitem[\protect\citeauthoryear{{Parise}, {Ceccarelli}, {Tielens}, {Castets}, {Caux}, {Lefloch}  \& {Maret}}{{Parise} et~al.}{2006}]{Parise2006}
{Parise} B.,  {Ceccarelli} C.,  {Tielens} A.~G.~G.~M.,  {Castets} A.,  {Caux} E.,  {Lefloch} B.,   {Maret} S.,  2006, \mn@doi [\aap] {10.1051/0004-6361:20054476}, \href {https://ui.adsabs.harvard.edu/abs/2006A&A...453..949P} {453, 949}

\bibitem[\protect\citeauthoryear{{Parise}, {Leurini}, {Schilke}, {Roueff}, {Thorwirth}  \& {Lis}}{{Parise} et~al.}{2009}]{Parise2009}
{Parise} B.,  {Leurini} S.,  {Schilke} P.,  {Roueff} E.,  {Thorwirth} S.,   {Lis} D.~C.,  2009, \mn@doi [\aap] {10.1051/0004-6361/200912774}, \href {https://ui.adsabs.harvard.edu/abs/2009A&A...508..737P} {508, 737}

\bibitem[\protect\citeauthoryear{{Pattle} et~al.,}{{Pattle} et~al.}{2019}]{Pattle2019}
{Pattle} K.,  et~al., 2019, \mn@doi [\apj] {10.3847/1538-4357/ab286f}, \href {https://ui.adsabs.harvard.edu/abs/2019ApJ...880...27P} {880, 27}

\bibitem[\protect\citeauthoryear{{Persson} et~al.,}{{Persson} et~al.}{2018}]{Persson2018}
{Persson} M.~V.,  et~al., 2018, \mn@doi [\aap] {10.1051/0004-6361/201731684}, \href {https://ui.adsabs.harvard.edu/abs/2018A&A...610A..54P} {610, A54}

\bibitem[\protect\citeauthoryear{{Podio} et~al.,}{{Podio} et~al.}{2024}]{Podio2024}
{Podio} L.,  et~al., 2024, \mn@doi [arXiv e-prints] {10.48550/arXiv.2407.04813}, \href {https://ui.adsabs.harvard.edu/abs/2024arXiv240704813P} {p. arXiv:2407.04813}

\bibitem[\protect\citeauthoryear{{Rawlings}, {Juvela}, {Lehtinen}, {Mattila}  \& {Lemke}}{{Rawlings} et~al.}{2013}]{Rawlings2013}
{Rawlings} M.~G.,  {Juvela} M.,  {Lehtinen} K.,  {Mattila} K.,   {Lemke} D.,  2013, \mn@doi [\mnras] {10.1093/mnras/sts233}, \href {https://ui.adsabs.harvard.edu/abs/2013MNRAS.428.2617R} {428, 2617}

\bibitem[\protect\citeauthoryear{{Roberts} \& {Millar}}{{Roberts} \& {Millar}}{2000}]{roberts2000}
{Roberts} H.,  {Millar} T.~J.,  2000, \aap, \href {https://ui.adsabs.harvard.edu/abs/2000A&A...364..780R} {364, 780}

\bibitem[\protect\citeauthoryear{{Roueff}, {Herbst}, {Lis}  \& {Phillips}}{{Roueff} et~al.}{2007}]{Roueff2007}
{Roueff} E.,  {Herbst} E.,  {Lis} D.~C.,   {Phillips} T.~G.,  2007, \mn@doi [\apjl] {10.1086/518863}, \href {https://ui.adsabs.harvard.edu/abs/2007ApJ...661L.159R} {661, L159}

\bibitem[\protect\citeauthoryear{{Roueff}, {Gerin}, {Lis}, {Wootten}, {Marcelino}, {Cernicharo}  \& {Tercero}}{{Roueff} et~al.}{2013}]{Roueff2013-CH2D+}
{Roueff} E.,  {Gerin} M.,  {Lis} D.~C.,  {Wootten} A.,  {Marcelino} N.,  {Cernicharo} J.,   {Tercero} B.,  2013, \mn@doi [Journal of Physical Chemistry A] {10.1021/jp400119a}, \href {https://ui.adsabs.harvard.edu/abs/2013JPCA..117.9959R} {117, 9959}

\bibitem[\protect\citeauthoryear{{Sabatini} et~al.,}{{Sabatini} et~al.}{2020}]{Sabatini2020}
{Sabatini} G.,  et~al., 2020, \mn@doi [\aap] {10.1051/0004-6361/202039010}, \href {https://ui.adsabs.harvard.edu/abs/2020A&A...644A..34S} {644, A34}

\bibitem[\protect\citeauthoryear{{Sabatini} et~al.,}{{Sabatini} et~al.}{2022}]{Sabatini2022}
{Sabatini} G.,  et~al., 2022, \mn@doi [\apj] {10.3847/1538-4357/ac83aa}, \href {https://ui.adsabs.harvard.edu/abs/2022ApJ...936...80S} {936, 80}

\bibitem[\protect\citeauthoryear{{Sabatini} et~al.,}{{Sabatini} et~al.}{2024}]{sabatini24}
{Sabatini} G.,  et~al., 2024, \mn@doi [\aap] {10.1051/0004-6361/202449616}, \href {https://ui.adsabs.harvard.edu/abs/2024A&A...684L..12S} {684, L12}

\bibitem[\protect\citeauthoryear{{Sakai} et~al.,}{{Sakai} et~al.}{2014a}]{Sakai2014a}
{Sakai} N.,  et~al., 2014a, \mn@doi [\nat] {10.1038/nature13000}, \href {http://adsabs.harvard.edu/abs/2014Natur.507...78S} {507, 78}

\bibitem[\protect\citeauthoryear{{Sakai} et~al.,}{{Sakai} et~al.}{2014b}]{Sakai2014b}
{Sakai} N.,  et~al., 2014b, \mn@doi [\apjl] {10.1088/2041-8205/791/2/L38}, \href {http://adsabs.harvard.edu/abs/2014ApJ...791L..38S} {791, L38}

\bibitem[\protect\citeauthoryear{{Santangelo}, {Murillo}, {Nisini}, {Codella}, {Bruderer}, {Lai}  \& {van Dishoeck}}{{Santangelo} et~al.}{2015}]{Santangelo2015}
{Santangelo} G.,  {Murillo} N.~M.,  {Nisini} B.,  {Codella} C.,  {Bruderer} S.,  {Lai} S.~P.,   {van Dishoeck} E.~F.,  2015, \mn@doi [\aap] {10.1051/0004-6361/201526428}, \href {https://ui.adsabs.harvard.edu/abs/2015A&A...581A..91S} {581, A91}

\bibitem[\protect\citeauthoryear{{Taquet} et~al.,}{{Taquet} et~al.}{2019}]{Taquet2019}
{Taquet} V.,  et~al., 2019, \mn@doi [\aap] {10.1051/0004-6361/201936044}, \href {https://ui.adsabs.harvard.edu/abs/2019A&A...632A..19T} {632, A19}

\bibitem[\protect\citeauthoryear{{Tielens}}{{Tielens}}{1983}]{Tielens1983}
{Tielens} A.~G.~G.~M.,  1983, \aap, \href {https://ui.adsabs.harvard.edu/abs/1983A&A...119..177T} {119, 177}

\bibitem[\protect\citeauthoryear{{Turner}}{{Turner}}{2001}]{Turner2001}
{Turner} B.~E.,  2001, \mn@doi [\apjs] {10.1086/322536}, \href {https://ui.adsabs.harvard.edu/abs/2001ApJS..136..579T} {136, 579}

\bibitem[\protect\citeauthoryear{{Vastel}, {Phillips}  \& {Yoshida}}{{Vastel} et~al.}{2004}]{vastel2004}
{Vastel} C.,  {Phillips} T.~G.,   {Yoshida} H.,  2004, \mn@doi [\apjl] {10.1086/421265}, \href {https://ui.adsabs.harvard.edu/abs/2004ApJ...606L.127V} {606, L127}

\bibitem[\protect\citeauthoryear{{Ward-Thompson}, {Kirk}, {Greaves}  \& {Andr{\'e}}}{{Ward-Thompson} et~al.}{2011}]{Ward2011}
{Ward-Thompson} D.,  {Kirk} J.~M.,  {Greaves} J.~S.,   {Andr{\'e}} P.,  2011, \mn@doi [\mnras] {10.1111/j.1365-2966.2011.18898.x}, \href {https://ui.adsabs.harvard.edu/abs/2011MNRAS.415.2812W} {415, 2812}

\bibitem[\protect\citeauthoryear{{van der Marel}, {Booth}, {Leemker}, {van Dishoeck}  \& {Ohashi}}{{van der Marel} et~al.}{2021}]{vandermarel21b}
{van der Marel} N.,  {Booth} A.~S.,  {Leemker} M.,  {van Dishoeck} E.~F.,   {Ohashi} S.,  2021, \mn@doi [\aap] {10.1051/0004-6361/202141051}, \href {https://ui.adsabs.harvard.edu/abs/2021A&A...651L...5V} {651, L5}

\makeatother
\end{thebibliography}




\clearpage 

\appendix

\section{Additional Material} 

\subsection{Integrated spectra towards 
VLA1623--2417}

\begin{figure}
\vspace{-1.5cm}
\includegraphics[width=9cm]{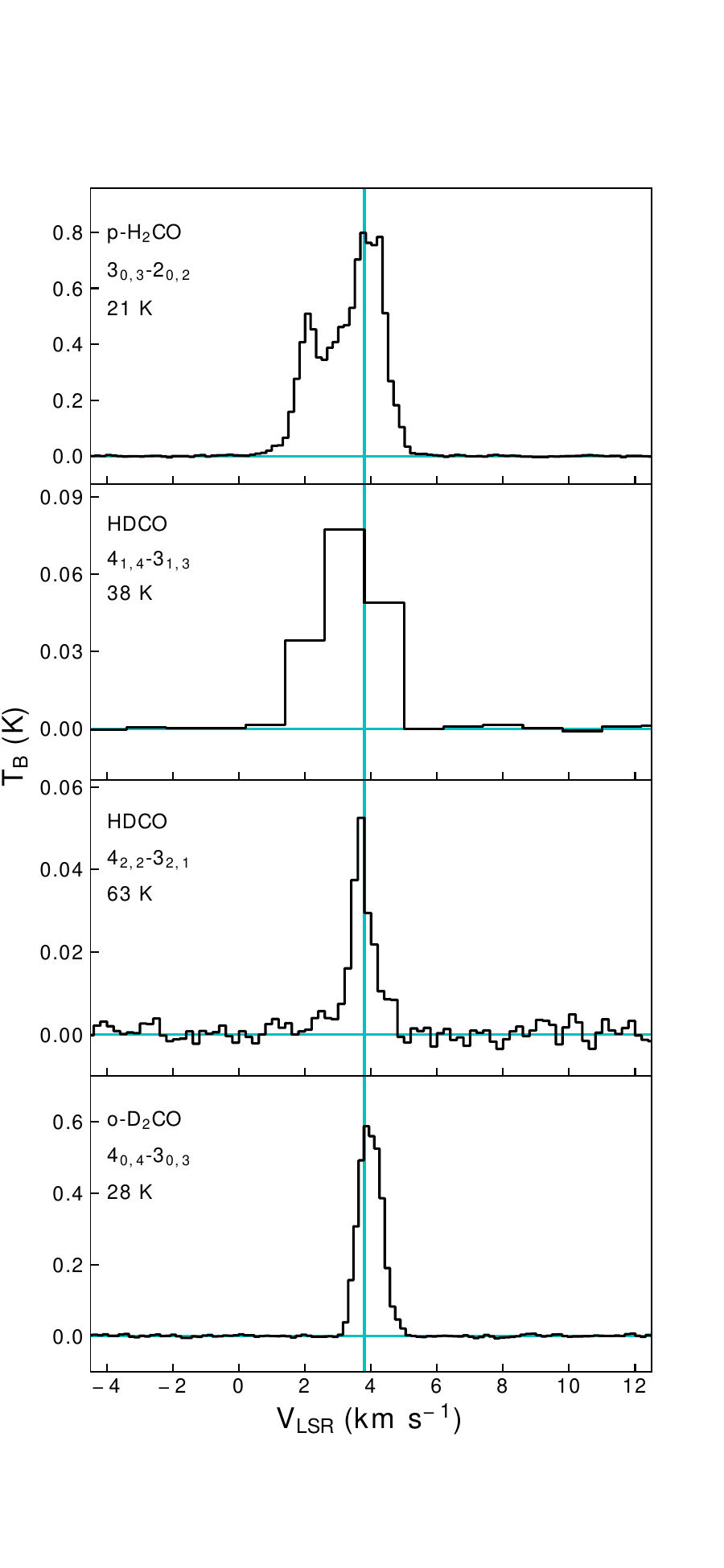}
\vspace{-1cm}
\caption{Spectra (in brightness temperature, $T_{\rm B}$, scale) observed towards VLA1623 and obtained integrating the emission over the FoV (see Table \ref{tab:lines}). {\it From top to bottom:} H$_2$CO (3$_{0,3}$ -- 2$_{0,2}$), HDCO (4$_{2,2}$ -- 3$_{2,1}$), HDCO (4$_{2,2}$ -- 3$_{2,1}$), and D$_2$CO (4$_{0,4}$ -- 3$_{0,3}$). The cyan vertical line represents the systemic velocity of VLA1623 A and B: $+3.8$ km s$^{-1}$ \citep{Ohashi2022}. 
}
\label{spectra}
\end{figure}

Figure \ref{spectra} reports the spectra (in brightness temperature, $T_{\rm B}$, scale) obtained integrating the line emission over the FoV of the  VLA1623--2417 observations (see Table \ref{tab:lines}).

\subsection{Channel maps of the HDCO lines}

\begin{figure*}
\includegraphics[width=14cm]{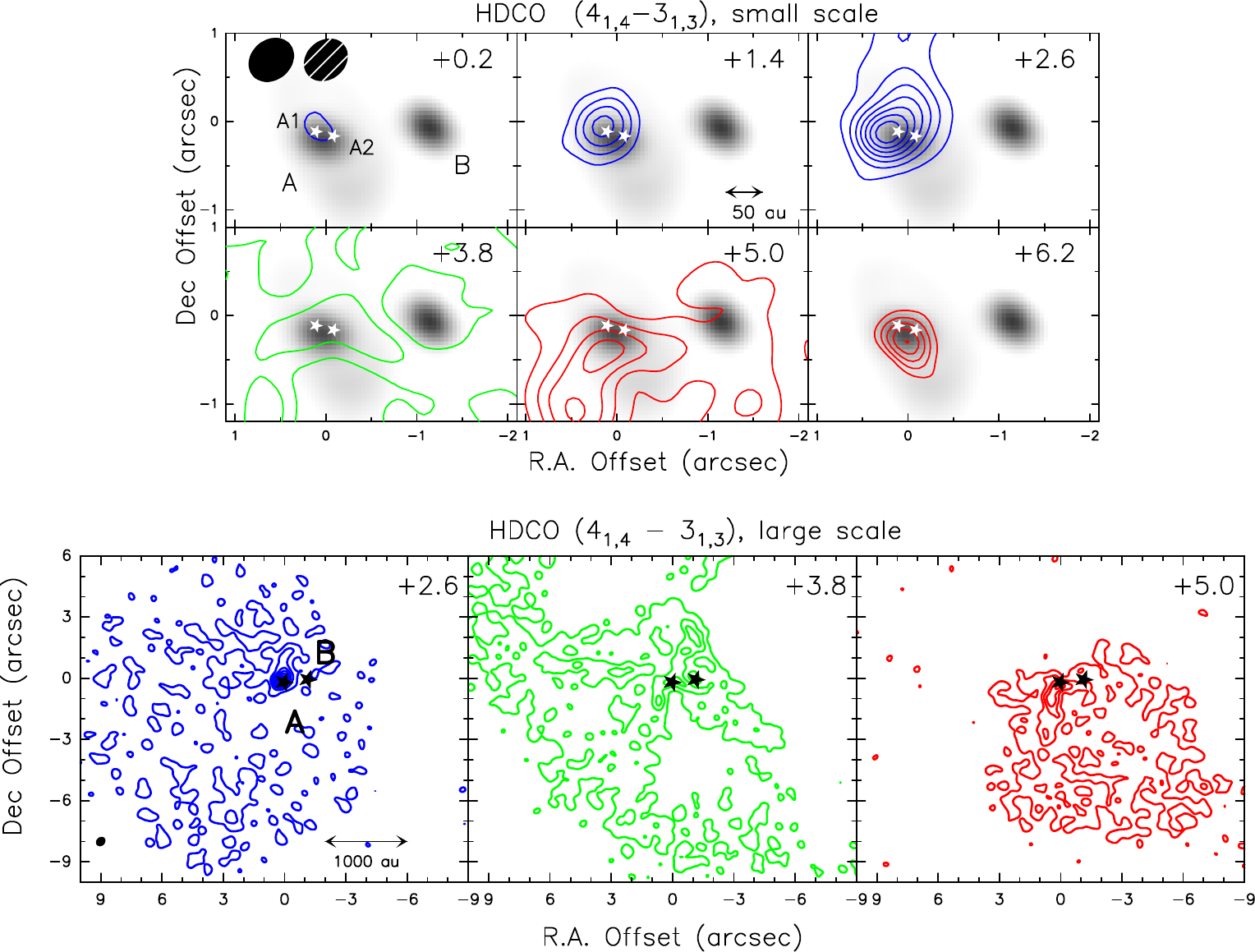}
 \caption{\textit{Top:} Channel maps of HDCO 4$_{1,4}$ -- 3$_{1,3}$ emission at small scale. The first contour and step is at 3$\sigma$ 
    (2.1 mJy beam$^{-1}$). The grayscale background shows the 1.3 mm continuum emission in Setup 2. The synthesized beams are shown in the top-left corner of the first channel by filled and dashed ellipses for line and continuum emission, respectively. White stars indicate the positions of VLA1623 A1 and A2 from \citep{Harris2018}. VLA 1623 A1, A2, A, and B are labeled in the first channel. \textit{Bottom:} Channel maps of HDCO 4$_{1,4}$ -- 3$_{1,3}$ emission at large scale, in the velocity range between +2.6 km s$^{-1}$ and +5.0 km s$^{-1}$. The black stars indicate VLA1623 A and VLA1623 B as detected in the FAUST continuum maps at 1.3~mm.
    }
 \label{HDCO_38K_APPENDIX}
\end{figure*}

\begin{figure*}
\includegraphics[width=16cm]{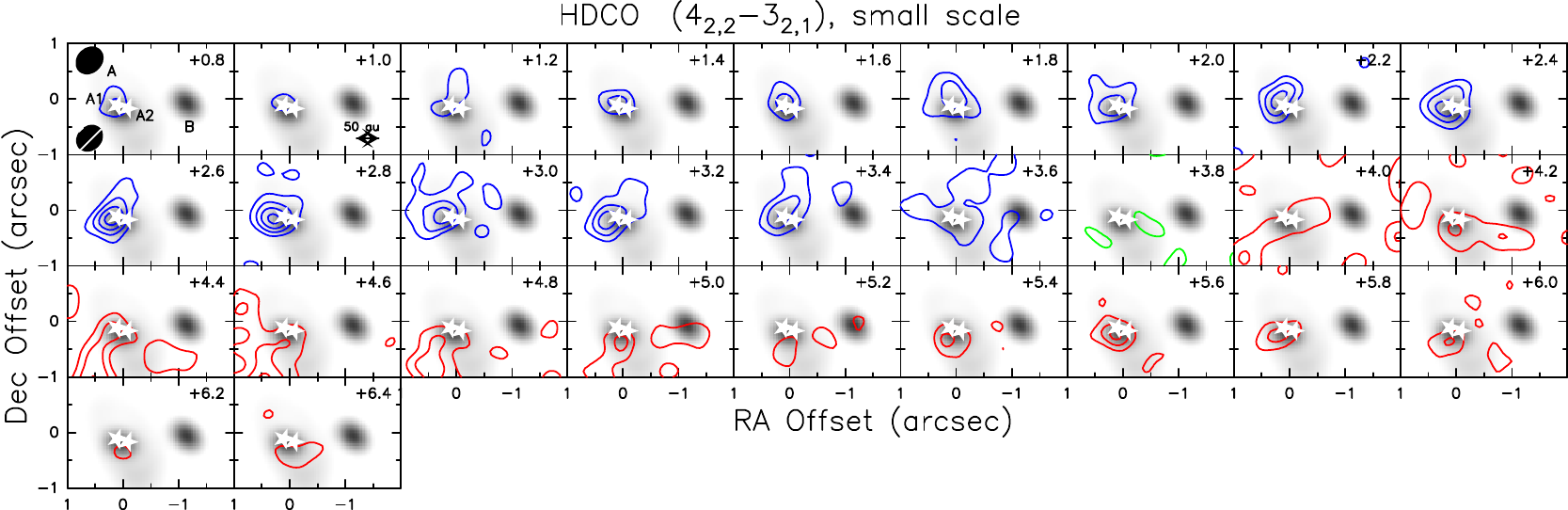}
 \includegraphics[width=14cm]{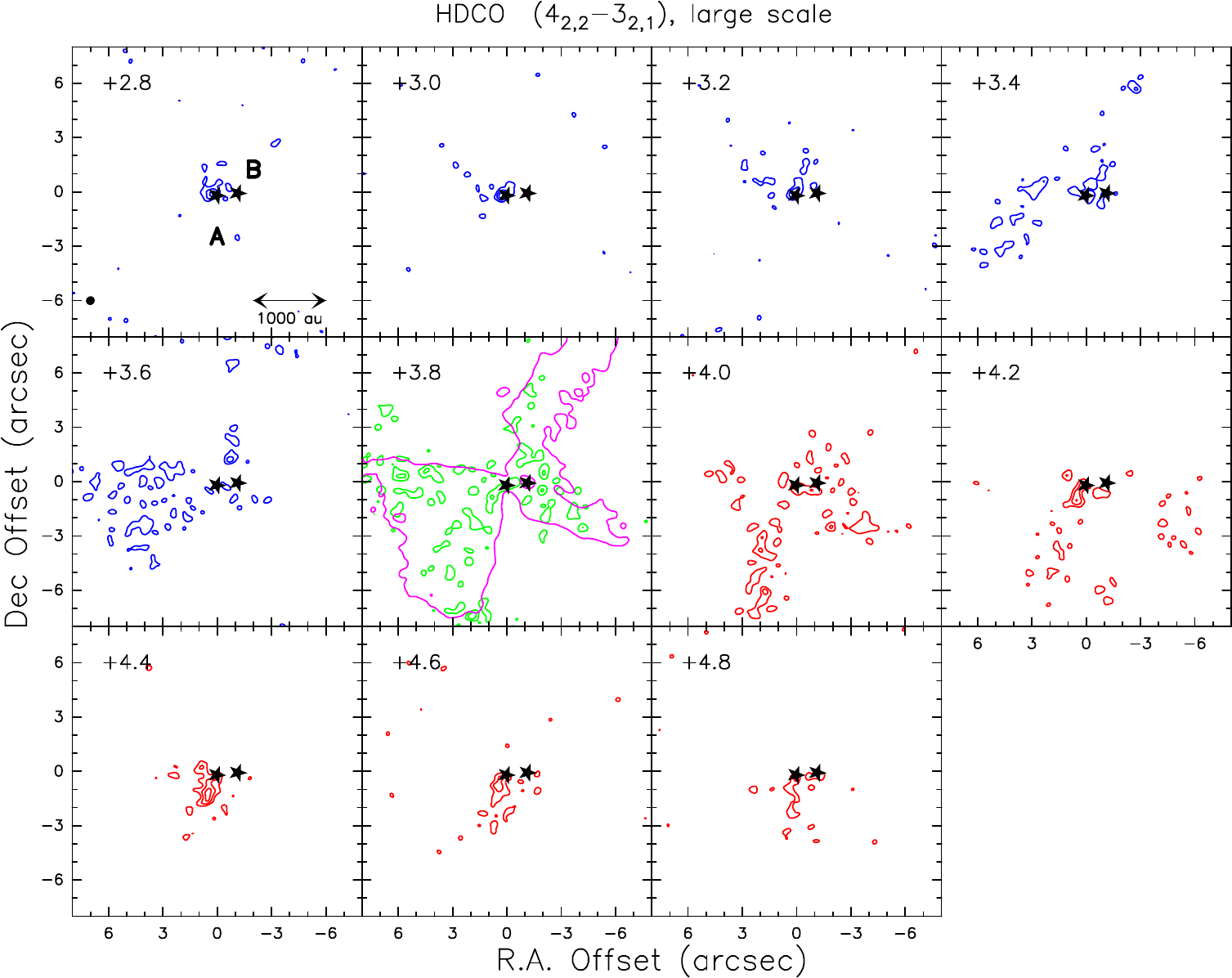}
 \caption{\textit{Top:} Channel maps of  HDCO 4$_{2,2}$ -- 3$_{2,1}$ at small scale. The first contour and step is at 6$\sigma$ (2.0 mJy beam$^{-1}$). The grayscale background shows the 1.3 mm continuum emission in Setup 2. The synthesized beams are shown in the left corner of the first channel by filled and dashed ellipses for line and continuum emission, respectively. White stars indicate the positions of VLA1623 A1 and A2 from \citep{Harris2018}. VLA1623 A1, A2, A, and B are labeled in the first channel. \textit{Bottom:} Channel maps of HDCO 4$_{1,4}$ -- 3$_{1,3}$ at large scale, in the velocity range ([+2.8, +4.8] km s$^{-1}$). The black stars indicate VLA1623 A and VLA1623 B as detected in the FAUST continuum map at 1.3~mm. The magenta contour in the channel at $+3.8$ km s$^{-1}$ indicates the outflow cavity walls probed by CS (5--4) \citep[25$\sigma$ contour, from][]{Ohashi2022}.}
    \label{HDCO_63K_APPENDIX}
\end{figure*}

We report the channel maps of the HDCO 4$_{1,4}$ -- 3$_{1,3}$ (Fig. \ref{HDCO_38K_APPENDIX}), and 4$_{2,2}$ -- 3$_{2,1}$ 
(Fig. \ref{HDCO_63K_APPENDIX}) emission. We show both high-velocity emission on small spatial scale (upper panels), and low-velocity emission on large spatial scales (lower panels).

\subsection{Integrated intensity (moment 0) maps of D$_{2}$CO (4$_{0,4}$ -- 3$_{0,3}$)}

\begin{figure*}
\centering
\includegraphics[width=18cm]
{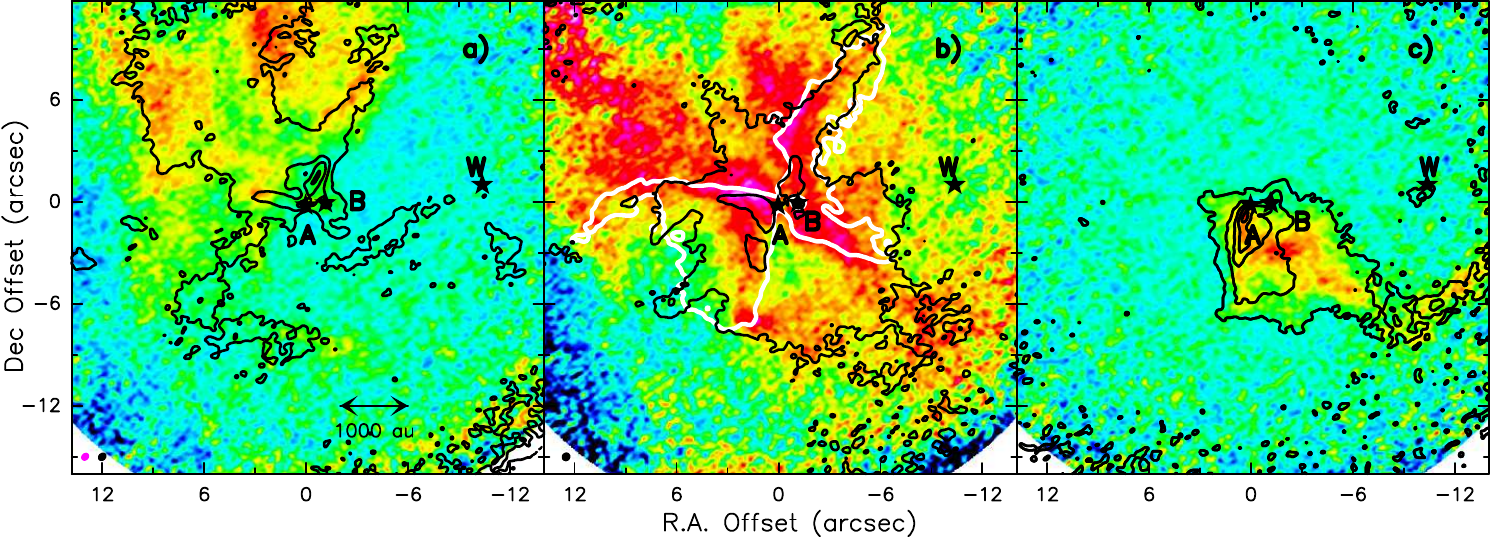}
    \caption{Integrated intensity (moment 0) maps of D$_2$CO (4$_{0,4}$ -- 3$_{0,3}$) emission (color) and H$_2$CO (3$_{0,3}$ -- 2$_{0,2}$) emission in black contours. The emission structures identified in the H$_{2}$CO and D$_{2}$CO emission in Panels a), b), and c) are labeled as V, VI, VII and are integrated on the velocity ranges listed in Table \ref{flows}. The first contour is at 3$\sigma$, and the step is 6$\sigma$: for Panel a) $\sigma$ is 2.2 mJy km s$^{-1}$ beam$^{-1}$, for Panel b) $\sigma$ is 12.0 mJy km s$^{-1}$ beam$^{-1}$, for Panel c) $\sigma$ is 4.5 mJy km s$^{-1}$ beam$^{-1}$. White contour in panel b) indicates the outflow cavity walls probed by CS (5--4) \citep[25$\sigma$ contour, from][]{Ohashi2022}. The synthesized beams are shown by the magenta and black ellipses in the bottom-left corner of the panel a) for the D$_{2}$CO and H$_{2}$CO, respectively, and by the black ellipse in the bottom-left corner of the panel b) for CS. The positions of VLA1623 A, B, and W are indicated by the black stars and are labeled. }
    \label{flows_D2CO_H2CO}
\end{figure*}

We present the integrated intensity maps (moment 0) of D$_2$CO (4$_{0,4}$ -- 3$_{0,3}$) towards VLA1623--2417 integrated on different velocity ranges in Fig. \ref{flows_D2CO_H2CO}.

\subsection{Channel maps of the D$_{2}$CO (4$_{2,3}$ -- 3$_{2,2}$) }
\begin{figure*}
\centering
\includegraphics[width=18cm]
{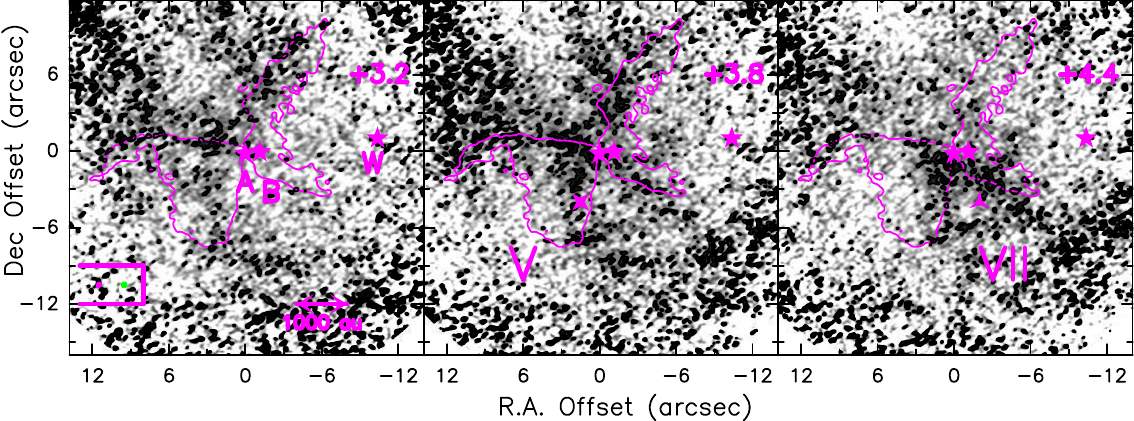}
    \caption{Channel maps of the D$_2$CO (4$_{2,3}$ -- 3$_{2,2}$) emission in the velocity range [+3.2, +4.4] km s$^{-1}$. The first contour is at 3$\sigma$, and the step is 3$\sigma$. The magenta contours indicate the outflow cavity walls probed by CS (5--4)  \citep[25$\sigma$ contour, from][]{Ohashi2022}. The magenta and green ellipses in the bottom-left corner of the first channel show the synthesized beams  for the CS (5--4) and D$_{2}$CO emission, respectively.
     The positions of VLA1623 A, B, and W are indicated by the magenta stars and are labeled in the first channel.
     The large-scale emissions are labeled as "VI" and "VII" in Table \ref{flows} consistently with the H$_2$CO components. Magenta cross and triangle symbols in the channels between +3.8 and +4.4 km s$^{-1}$ show the position along the outflow cavity wall and the streamer where the spectra are extracted (Fig. \ref{superspectra} in Sect. \ref{D2}).}
    \label{channel_D2CO_50K}
\end{figure*}

Figure \ref{channel_D2CO_50K} shows the channel maps of the D$_{2}$CO (4$_{2,3}$ -- 3$_{2,2}$) emission in VLA1623-2417. 

\subsection{Intensity ratio maps of HDCO/H$_2$CO and D$_2$CO/H$_2$CO}
\begin{figure*}
\centering
\includegraphics[width=18cm]{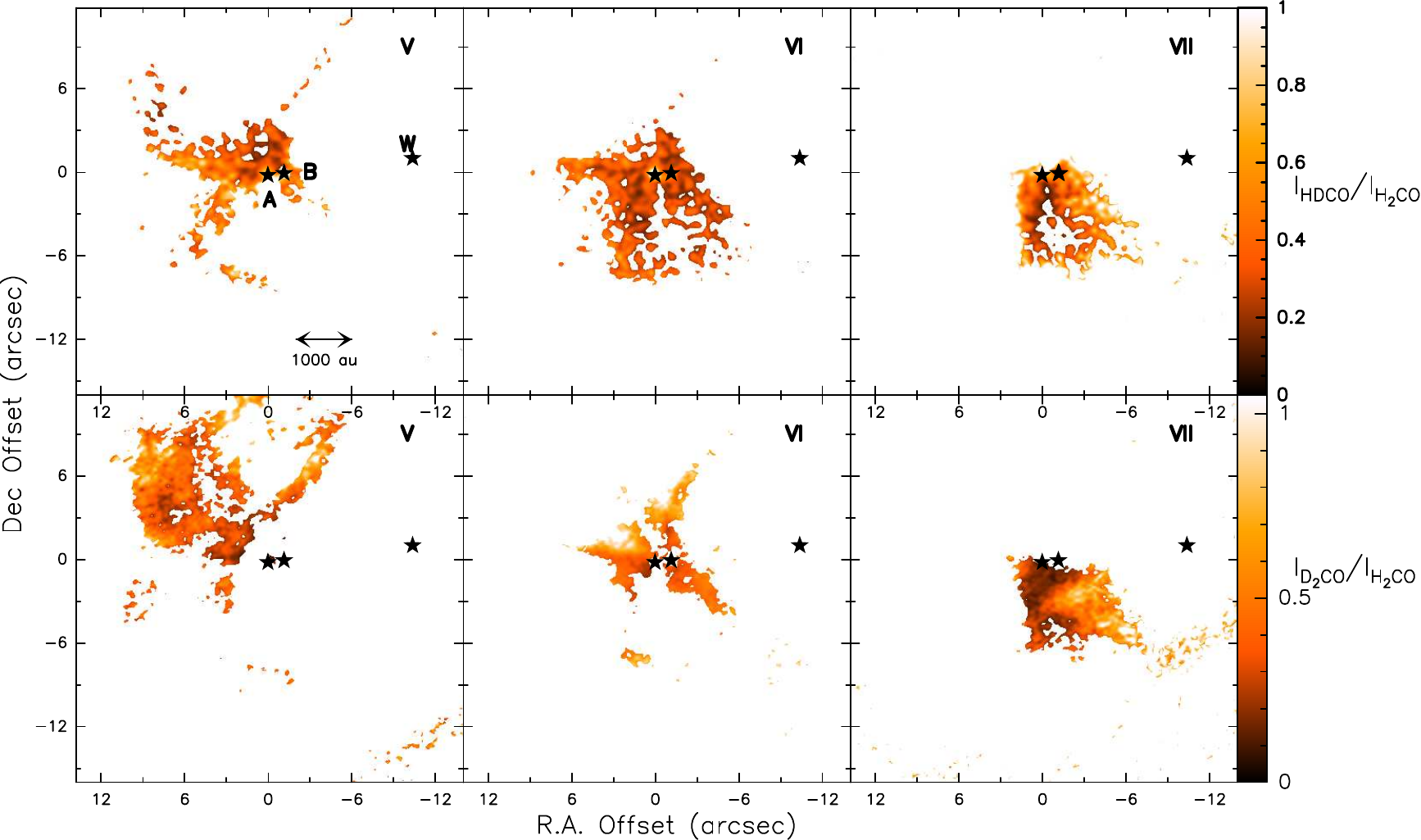}
    \caption{Intensity ratio maps of HDCO(4$_{1,4}$ -- 3$_{1,3}$)/p-H$_2$CO(3$_{0,3}$ -- 2$_{0,2}$)  (upper rows) and o-D$_2$CO(4$_{0,4}$ -- 3$_{0,3}$)/p-H$_2$CO(3$_{0,3}$ -- 2$_{0,2}$)  (bottom rows) for three components of the  VLA1623-2417 region (see Tab. \ref{flows}): northern blue-shifted streamer towards A (label V), outflow cavities driven by A (VI), and southern red-shifted streamer towards A (VII). The color bars are shown at the right corners related to the ratio molecules. The positions of VLA1623 A, B, and W are indicated by the black stars, and labeled in the first panel.}
    \label{RatioMaps}
\end{figure*}

Figure \ref{RatioMaps} shows the Intensity ratio maps of HDCO(4$_{1,4}$ -- 3$_{1,3}$)/p-H$_2$CO(3$_{0,3}$ -- 2$_{0,2}$)  (upper rows) and o-D$_2$CO(4$_{0,4}$ -- 3$_{0,3}$)/p-H$_2$CO(3$_{0,3}$ -- 2$_{0,2}$)  (bottom rows) towards the VLA1623-2417. region. In particular, three components are imaged:
(see Tab. \ref{flows}): northern blue-shifted streamer towards A (label V), outflow cavities driven by A (VI), and southern red-shifted streamer towards A (VII).

\subsection{Intensity-weighted mean velocity and velocity-dispersion maps of D$_{2}$CO (4$_{0,4}$ -- 3$_{0,3}$)}

\begin{figure*}
\includegraphics[width=8cm]{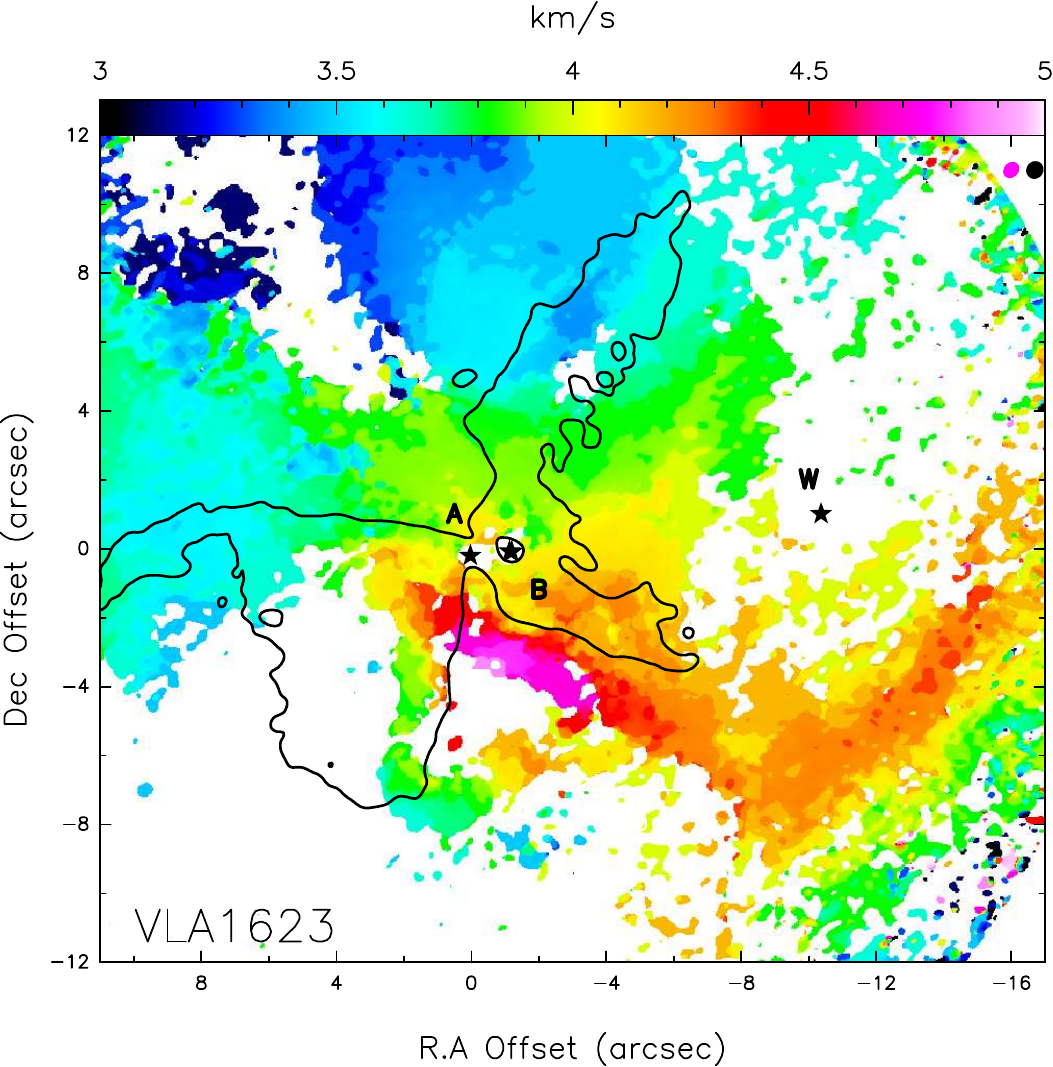}
\includegraphics[width=8cm]{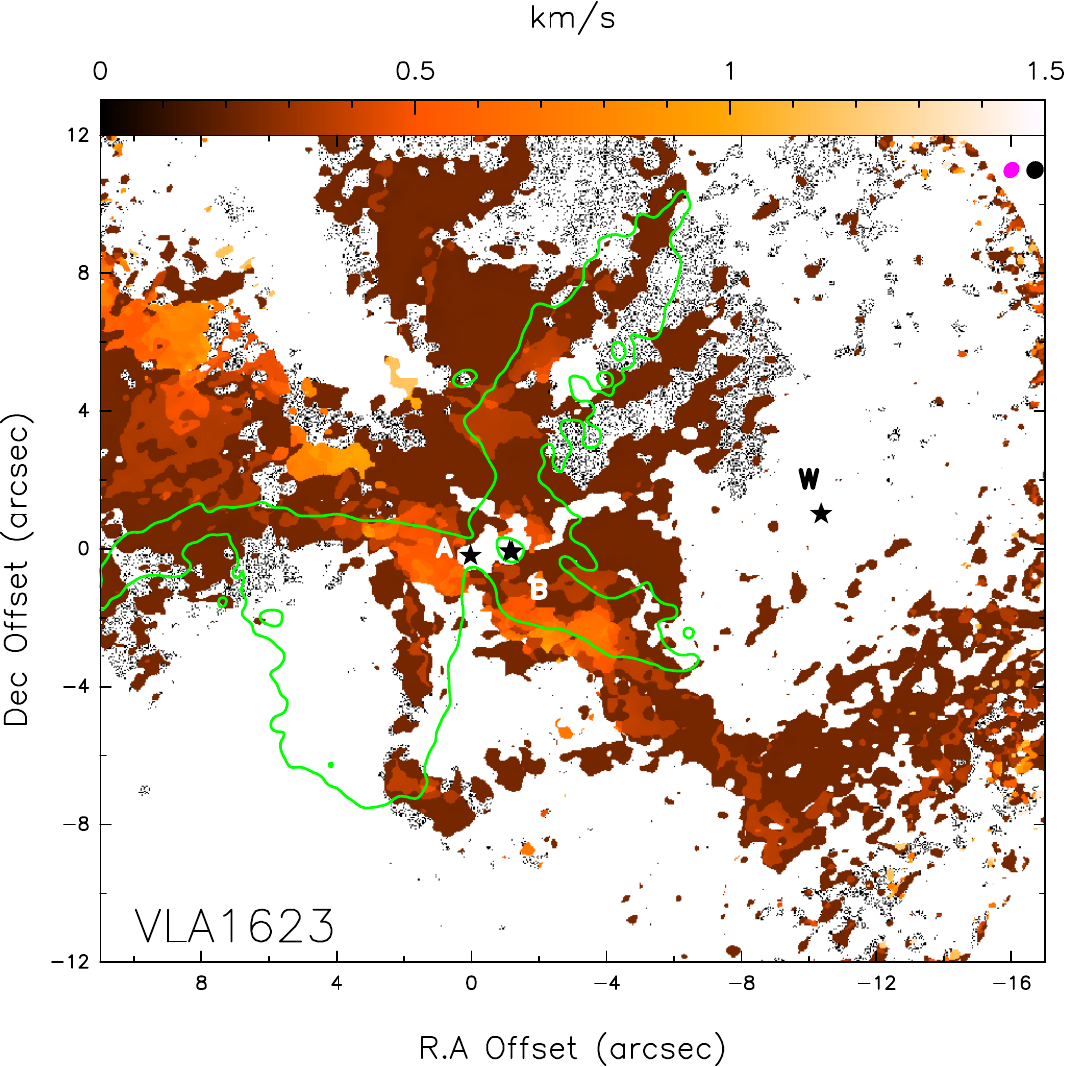}
    \caption{{\it Left:} Intensity-weighted mean velocity (moment 1) map of D$_2$CO (4$_{0,4}$ -- 3$_{0,3}$) emission towards VLA1623--2417 over the velocity range between +3.0 km s$^{-1}$ and +5.2 km s$^{-1}$. The black contour indicates the outflow cavity walls probed by CS (5--4) \citep[25$\sigma$ contour, from][]{Ohashi2022}.
    The synthesized beams are shown by the magenta and black ellipses in the top-right corner for the D$_{2}$CO and CS (5--4) emission, respectively (see Tab. \ref{tab:lines}). {\it Right:} Velocity–dispersion (moment 2) map of D$_2$CO (4$_{0,4}$ -- 3$_{0,3}$) towards VLA1623--2417 over the velocity range between +3.0 km s$^{-1}$ and +5.2 km s$^{-1}$. The green contour indicates the outflow cavity walls probed by CS (5--4). The positions of VLA1623 A, B, and W are indicated by the black stars, and labeled in the first panel.}
    \label{D2CO_Mom1and2}
\end{figure*}

We present the intensity-weighted mean velocity (moment 1) and velocity–dispersion (moment 2) maps of D$_2$CO (4$_{0,4}$ -- 3$_{0,3}$) emission spanning the velocity range from +3.0 km s$^{-1}$ to +5.2 km s$^{-1}$ towards VLA1623--2417 in Fig. \ref{D2CO_Mom1and2}


\bsp	
\label{lastpage}
\end{document}